\documentclass[journal=jpcbfk,manuscript=article]{achemso}
\setkeys{acs}{super = true}
\setkeys{acs}{articletitle=true}
\setkeys{acs}{maxauthors=10}
\setkeys{acs}{etalmode=truncate}
\setcitestyle{super,open={},close={}}

\usepackage{array}
\usepackage{dcolumn}
\usepackage{graphicx}
\usepackage{xspace}
\xspaceaddexceptions{]\}}
\usepackage{textcomp}
\usepackage{lineno}
\usepackage{color}
\usepackage{lscape}
\usepackage{amsmath}
\usepackage[version=4]{mhchem} 
\SectionNumbersOn
\AtBeginDocument{\nocite{achemso-control}}
\newcommand{\chem}[1]{\ensuremath{\mathrm{#1}}}

\newcommand {\kcal} {kcal mol$^{-1}$\xspace}
\newcommand {\mcal} {Mcal\xspace}

\newcommand {\xx}{$\mathbf{X}^{+Q}$}
\newcommand {\yy}{$\mathbf{Y}^{-Q}$}
\newcommand {\allatom}{\emph{all atom}\xspace}
\newcommand {\ie}{\emph{i.e.}\xspace}
\newcommand {\ppp}{\textit{ppp}\xspace}
\newcommand {\pppl}{\textit{ppp}$^l$\xspace}
\newcommand {\pppa}{\textit{ppp}$^2$\xspace}

\newcommand {\cl}{\ce{Cl^-}\xspace}
\newcommand {\na}{\ce{Na^+}\xspace}
\newcommand {\xb}{$\mathbf{X}^{2+}$\xspace}
\newcommand {\yb}{$\mathbf{Y}^{2-}$\xspace}

\newcommand{\hpp}{\textbf{HPP}\xspace}

\newcommand{\va}{V \AA$^{-1}$\xspace}

\title{Modeling polyelectrolyte hydration from a multi scale polarizable pseudo particle solvent coarse grained approach.}

\author{Michel Masella}
\affiliation{Laboratoire de Biologie Structurale et Radiobiologie, Service de Bio\'energ\'etique, Biologie Structurale et M\'ecanismes, 
    Institut de Biologie et de Technologies de Saclay, CEA Saclay, F-91191 Gif sur Yvette Cedex, France}
\email{michel.masella@cea.fr}

\author{Fabien L\'eonfort\'e}
\affiliation{L'Or\'eal Group, Research \& Innovation, Aulnay-Sous-Bois,France}

\begin{document}

\maketitle

\newpage

\begin{abstract}

We investigate the reliability of simulations of polyelectrolyte systems in aqueous environments, simulations that are performed using an efficient multi scale coarse grained polarizable pseudo-particle particle approach, denoted as \pppl, to model the solvent water, whereas the solutes are modeled using a polarizable \allatom force field. We focus our study on issues tied to two key parameters of the \pppl approach, namely the extension of the solvent domain SD at the close vicinity of a solute (domain in which each solvent particle corresponds to a single water molecule) and the magnitude of solute/solvent short range polarization damping effects. To this end we built a new \pppl models from which we simulate NaCl aqueous solutions at the molar concentration scale. We also  re investigate the hydration of a hydrophobic polyelectrolyte polymer that we showed in an earlier study [J Chem Phys, 114903 (155) 2021] to evolve towards a counter intuitive globular  form surrounded by a spherical counter ion cloud along \pppl-based simulations. Strong short range damping is pivotal to simulate NaCl aqueous solutions.  The extension of the domain SD (as well as short range damping) has a weak effect on the conformation of the polymer, but it plays a pivotal role to compute accurate solute/solvent interaction energies. In all our results lead us to recommend to simulate polyelectrolyte polymers as dissolved alone in \pppl fluids (\ie without explicitly accounting for their counter ions) to investigate their behavior at infinite dilution conditions, and to systematically consider strong solute/solvent polarization short range damping to model charged species.
 
\end{abstract}

\small {\noindent \textbf{Keywords} Polarization. Coarse grained model. Salt solutions. Polyelectrolyte polymer.}

\newpage

\section{Introduction}

Coarse Grained, CG, approaches are efficient molecular modeling scheme, which are more and more commonly used to investigate the properties of very large molecular systems that are beyond the computational capacity of standard \allatom force fields (\ie interaction potentials that explicitly take into account all the atoms of a molecular system). We may cite among others (see the recent reviews Refs. \cite{kmiecik16,joshi21,noid23}) the MARTINI force field to model lipids, proteins, carbohydrates and other biomolecules \cite{marrink23}, the CG approach devoted to study complex aggregates made of huge chitosan chains (at the 1 000 unit size scale) in salty aqueous solutions  \cite{tsereteli17}, as well as the water CG approach mW \cite{molinero09}.  CG approaches are of particular interest  in the industrial research and development field (from chemistry, oil manufacturers to personal care/cosmetics) \cite{ingo14,hospital15,lu16,illa18,mehand20,nakamura20} because of the complexity of most of the solutions that these industries use and the size of those solutions ingredients, like chitosan chains. Most of CG approaches rely on simple pair wise interaction potentials whose parameters are assigned from macroscopic properties (top down schemes, see for instance \citet{chen18}) and/or from microscopic features of a molecular systems (bottom up schemes, see among others \citet{shell08}) using different numerical methods, like force matching \cite {wang09,bacle20} and  machine learning  \cite{behler16,zhang18,wang19,li20} techniques. Recently sophisticated CG approaches accounting for many-body effects (by approximating the total potential energy of a molecular system as a sum $N$-body energy components) have emerged \cite{john17,wang21}.
 
In a series of articles \cite{masella08,masella11,masella13} we detailed a multi-scale version, denoted as \pppl, of the original polarizable hybrid CG scheme of \citet{haduong02} that was built to simulate the hydration of explicit solutes modeled using \allatom force fields and as dissolved in a fluid of polarizable pseudo particles (denoted as \ppp's). Such kind of CG scheme belongs to the wide class of solvent implicit continuum approaches, like the popular Polarizable Continuum Model, PCM \cite{tomsai05}. However, as the original \ppp approach, its multi scale \pppl version retains the notion of particle as it relies on a hierarchical representation of a solvent  surrounding a solute: at short range from that solute the solvent is modeled by polarizable pseudo particles whose size corresponds to that of a single solvent molecule, whereas at longer and longer range from the solute larger and larger local volumes of the solvent are modeled by means of larger and larger \pppl polarizable particles. Despite its sophistication, the computational complexity of the \pppl approach scales as $O(N)$ and it allows one to readily simulate using reduced computational ressources  the hydration of complex solutes in extended water environments, as we recently showed for medium-size chitosan chains  \cite{masella23} as well as for a hydrophobic polyelectrolyte polymer dissolved in an aqueous environment comprising the equivalent of 10 M water molecules \cite{masella21}.

To model water and regardless of their size,  the (isotropic) polarizability $\alpha_s$ of the \pppl particles obeys the Clausius-Mosotti relation \cite{masella08,masella11} 

\begin{equation} \label{eqn:ppp_polarizability}
\alpha_s = \frac{\epsilon_s -1}{4\pi \rho_s \epsilon_s}. 
\end{equation}
Here  $\epsilon_s $ and $\rho_s$ are the dielectric constant and the density of liquid water, respectively. The natural choice is to set the particle volume to that of a water molecular. However the above relation allows one to consider a particle size that corresponds to a solvent local volume comprising a cluster of water molecules. We used that feature to propose the multi-scale version \pppl  depicted in Figure \ref{fig:multiscale}. 

By setting the particle volume to that of a single water molecule, relation (\ref{eqn:ppp_polarizability}) yields $\alpha_s = 2.35$ \AA$^3$, a value that is 60\% larger than the isotropic polarizability of a water molecule (1.45 \AA$^3$). That shows $\alpha_s$ to not correspond to a standard microscopic (atomic or  molecular) polarizability:  it also allows to account for solvent molecular orientational polarization (\ie water orientational perturbation arising from the solute presence) \cite{haduong02}.  

A second important assumption of our \pppl approach is the neglect of intra-solvent polarization effects: the magnitude of the particle induced dipole moments $\mathbf{p}_s$ is only a function of the solute electric field $\mathbf{E}_\mathrm{solute}$ acting on the particle centers. However that  '\textit{local}' approximation may yield solute/solvent over polarization phenomena. To circumvent them we follow the original ideas of \citet{haduong02} by allowing the dipoles of the smallest particles (those which interact at short range from a solute) to saturate according to

\begin{equation} \label{eqn:langevin_solvent}
\mathbf{p}_s= \mu_s \, {\cal{L}} \left( \frac{3 \alpha_s \mathbf{E}_\mathrm{solute} }{\mu_s} \right) \frac{\mathbf{E}_\mathrm{solute} }{\left| \mathbf{E}_\mathrm{solute}\right|} ,
\end{equation}
here $\cal{L}$ is the Langevin function and $\mu_s$ is the particle saturation dipole value. The corresponding solvent/solute polarization energy for a system comprising $N_s$ particles is then

\begin{equation} \label{eqn:ppp_polarization}
U^{pol}_{ps} = - \frac{\mu_s^2}{3 \alpha_s} \sum_{j=1}^{N_s} \ln \left[ \frac{\sinh \left(3\alpha_s  \left| \mathbf{E}_\mathrm{solute}^j \right| / \mu_s \right)}{3\alpha_s \left| \mathbf{E}_\mathrm{solute}^j \right| \mu_s} \right].
\end{equation}

All modern \allatom polarizable force fields (\ie interaction potentials that model water as an explicit  tri-atomic polarizable molecule) account for short range damping of the electric fields \cite{real13,lemkul16,jing19}. As the \ppp particles are expected to behave as real water molecules at short range from a solute, a reasonable assumption is also to damp the solute electric fields  $\mathbf{E}_\mathrm{solute}$ acting on the particles at short range. As they both weaken the solute electric field, dipole saturation and electric field short range damping are redundant effects, and it is far from obvious to disentangle them in order to build a transferable CG approach able to accurately model complex charged microscopic systems (from salty aqueous solutions, ionic liquids up to macro polyelectrolyte polymers) under different chemical conditions.

The aim of the present study is to discuss particular features of the \pppl approach (like the role of dipole saturation and solute/solvent electric field short range damping) and to further discuss its reliability to simulate complex and highly charged solutes. To this end we investigate  in the aqueous phase by means of the \pppl approach the effects of the presence of an opposite charge ion pair (whose cation/anion absolute charge varies from +1 to +4 $e$) on water at medium range, as well as ion association in NaCl solutions whose salt concentrations span from 0.2 to 1.0M (and for different intensities of the solute electric field short range damping). To assess the reliability of the \pppl approach  we compare \pppl simulation results to available data from accurate  \allatom polarizable force fields  \cite{soniat16,vallet22} and/or to new \allatom simulations. 

We also re investigate the structural behavior of a hydrophobic polyelectrolyte polymer in the aqueous phase using a new \pppl model that corresponds to a stronger solute/solvent electric field damping than in our original study \cite{masella21}. In the latter study, we showed that polymer (simulated with its counter ions) to rapidly collapse towards a globular form. Even if the goal of that original study was to discuss the strength of dynamic microscopic polarization effects on complex polyelectrolyte polymers, the nature of such a counter intuitive globular form for a heavily charged polymer needs to be further investigated to assess the reliability of the approach \pppl. However first we will shortly detail the \pppl approach and the main features of its latest version. Below we label as '\ppp' and '\pppl' the CG approaches, simulations and data generated using only small particles (whose volume is that of real water molecule) and set of particles of different size, respectively.

\section{Computational details}

\subsection{Long range interaction truncation scheme}

Solute/\ppp particle electrostatic (polarization) interactions are truncated according to a shell-based cut off scheme: a \ppp particle undergoes the electric field from all the solute electrostatic charges if the smallest distance among that particle and all the non-hydrogen solute atoms is smaller than a reference cut off distance $R_{cut}$. For other kinds of short range pair interactions, truncation is achieved using a standard spherical radius-based cut off scheme. 

Interaction truncation is performed by scaling a pair potential energy component by means of a function $G$ that smoothly downscales it for distances that span between $R_{cut}$ and $R_{cut}+ \delta R$. We systematically set $\delta R$ to 0.5 \AA $ $ and the function $G$ is defined from a fifth order B-spline polyn\^ome $P_5$ according to 

\begin{equation} \label{eqn:cutoff}
G(r,R_{cut},\delta R) = \left\{ \begin{array}{lcc}
  1 &\text{if}& r \leq R_{cut} \\
  0 &\text{if}& r \geq R_{cut} + \delta R \\
  P_5(r) &\text{if}& R_{cut} < r < R_{cut} + \delta R \\
\end{array} \right.
\end{equation}

\subsection{The \ppp approach} \label{sec:cg}

The  electric field generated by an atom $j$ on a \ppp particle $i$ in Equations (\ref{eqn:langevin_solvent}) and (\ref{eqn:ppp_polarization}) obeys

\begin{equation} \label{eqn:electric_field}
\textbf{E}_\mathrm{solute}^{j}=\frac{1}{4\pi\epsilon_{0}r_{ij}^{3}}  \left[
 \tilde{q}_{i}(\mathbf{r}_{i}-\mathbf{r}_{j}) - \tilde{T}{\vert \mathbf{r}_{i}-\mathbf{r}_{j}\vert} \right],
\end{equation}
$\mathbf{r}_i$, $\mathbf{r}_j$ and $r_{ij}$ are the vector positions and the distance between the centers $i$ and $j$. The functions $\tilde{q}_i$ and $\tilde{T}{\vert \mathbf{r}_{i}-\mathbf{r}_{j}\vert}$ are:

\begin{equation}
\tilde{q}_i = \lambda_{3,i} q_i \:\: \text {     and     } \:\:
\tilde{T}{\vert \mathbf{r}_{i}-\mathbf{r}_{j}\vert} = 3\lambda_{5,i}\frac{(\mathbf{p}_i \cdot \mathbf{r}_{i}) \times \mathbf{r}_{j}}{{\vert \mathbf{r}_{i}-\mathbf{r}_{j}\vert}^{2}}-\lambda_{3,i} \mathbf{p}_i,
\end{equation}
here, $q_i$ and $\mathbf{p}_i$ are the static Coulombic charge and the induced dipole moment of atom $i$. $\lambda_{3,i}$ and $\lambda_{5,i}$ are two functions that monitors the magnitude of the electric field damping at short range. 

On the other hand an atom $i$ undergoes the electric field $\textbf{E}_\mathrm{ppp}^{i}$ generated by the induced dipoles of the \ppp particles. That electric field has the same analytical form as $\textbf{E}_\mathrm{solute}^{j}$ by setting $\tilde{q}_j \equiv 0 $. We set  the saturation parameter $\mu_s$ for atom dipole/atom dipole interactions to a large value (12 Debye), so that the standard dipole/dipole interaction potential is recovered in that case. 

Based on the original work of Thole \cite{thole81}, we consider as short range damping functions:

\begin{equation}
\lambda_{3,i}=1-\exp\left( {-a_i \times {r_{ij}^{3}}}\right) \:\: \text{    and    } \:\:
\lambda_{5,i}=1-(1+a_i \times {r_{ij}^{3}})\exp\left( {-a_i \times {r_{ij}^{3}}}\right) .
\end{equation}
$a_i$ is a parameter, expressed in \AA$^{-3}$, which depends only on the nature of atom $i$. The weaker is $a_i$ the stronger are the corresponding damping effects, which are truncated for distances $r_{ij}$ larger than 5.0 \AA. 

Atom/\ppp interactions are truncated for distances $r_{ij}$ larger than $R_\mathrm{cut,1}^{pol} = 12$ \AA. Assuming the \ppp density to be constant at long range from a non polarizable point charge $Q$, the long range electrostatic energy not taken into account because of truncation is

\begin{equation} \label{eqn:precision}
\delta G_\mathrm{lr} =- \frac{\alpha_s \rho_s}{4\pi \epsilon_0} \int_{R_\mathrm{cut,1}^{pol}}^\infty \left( \frac{Q}{r^2} \right)^2  4\pi r^2 dr = - \frac{Q^2}{8\pi \epsilon_0 R_\mathrm{cut,1}^{pol}}.
\end{equation}
Accounting for the smoothing function $G$, that relation yields $\delta G_\mathrm{lr} = -13.5$ \kcal for a $R_\mathrm{cut,1}^{pol} = 12$ \AA $ $ and $|Q| = 1$ $e$. For complex non-symmetric solutes, we proposed a multi-scale approach to compensate truncation, see Section \ref{sec:long_range} below.

Besides solute/solvent polarization, the \ppp model also accounts for solute/solvent non-electrostatic interactions by means of a  Lennard-Jones-like potential truncated for atom/\ppp distances larger than $R_\mathrm{cut,1}^{pol}$:
 
\begin{equation}
\tilde{U}^{LJ} = \sum_{i=1}^{N_a} \sum_{j=1}^{N_s}  \epsilon_{i,\ppp}^* \left[ \left(\frac{\sigma_{i,\ppp}^*}{r_{ij}} \right)^{6m} - m \left( \frac{\sigma_{i,\ppp}^*}{r_{ij}} \right)^{6} \right] \times G(r_{ij}).
\end{equation}
$N_a$ and $N_s$ are the numbers of atoms and \ppp's, respectively. As in our original study  \cite{masella13}, we set $m$ to 3. For given $\mu_s$ and $a_i$ values, the  parameters $\sigma_{i,\ppp}$ and $\epsilon_{i,\ppp}$ are assigned to best reproduce (within 0.1 \kcal) the available experimental hydration Gibbs free energy for a set a target entities, like ions \na and \cl, as well as the mean solute/water distances in the entity first hydration shell. We don't pay attention in the present study to reproduce ion first shell coordination numbers.
  
Interactions among \emph{ppp}'s are also modeled using Lennard-Jones-like  energy term $\tilde{U}^{LJ}_\ppp$ and a many-body  term $U^{dens}$ that is a function of the measures $n_0$ and $n_1$ of the instantaneous coordination numbers in first and second hydration shells of a \emph{ppp}, which are estimated as in our original study \cite{masella11}. However we now slightly alter the original term $U^{dens}$ as

\begin{equation}
U^{dens} = \sum_{j=1}^{N_s} \sum_{p=1,2} \nu_p \times \max \left[ (n_p - \bar{n}_p,0) \right]^2.
\end{equation} 
 
We tested several different $\tilde{U}^{LJ}_\ppp$ potentials in conjunction with $U^{dens}$. Our attempts show the  choice of $\tilde{U}^{LJ}_\ppp$ to have negligible effects on simulation results. All the data discussed below are based on:
 
\begin{equation}
 \tilde{U}^{LJ}_\ppp = \sum_{j=1}^{N_s}  \sum_{k=j+1}^{N_s}  \epsilon_{\ppp}^* \left[ \left(\frac{\sigma_{\ppp}^*}{r_{jk}} \right)^{12} - 4 \left( \frac{\sigma_{\ppp}^*}{r_{jk}} \right)^{3} \right].
\end{equation}  
The parameters of $\tilde{U}^{LJ}_\ppp$ and $U^{dens}$ (namely $\epsilon^*_\ppp$, $\sigma^*_\ppp$, and the $\nu_p$'s and $\bar{n}_p$'s) are assigned to reproduce (at ambient conditions) the hydration Gibbs  free energy of a water molecule (6.3 \kcal), the water density (0.0335 molecules per \AA$^3$) and the two regimes of the free energy corresponding to the creation of an empty cavity in neat water  according to the Lum-Chandler-Weeks theory of hydrophobicity \cite{chandler05}. The latter conditions are met by truncating the terms $\tilde{U}^{LJ}_\ppp$ and $U^{dens}$ for inter \ppp distances larger than $R^{cut}_\ppp$ = 7 \AA, whereas a larger cut off distance is needed (at least about 15 \AA) when only taking into account $\tilde{U}^{LJ}_\ppp$. As the computational time to estimate interactions  among \ppp's scales as $\left({R^{cut}_\ppp}\right)^3$, the use of the term $U^{dens}$ provides an efficient way to compute them.
 
 In the present implementation of the \ppp model, we use cubic periodic conditions to maintain constant the mean  \ppp density along NPT simulations, on average. Because of the truncation of long range inter \ppp interactions, we usually need to only account for the 27 first \ppp periodic images. Moreover  solute atoms do not interact with their periodic images, which allows us to simulate a solute at the infinite dilution conditions. For a single molecule system (even a large protein or a polymer) the latter approach does not introduce artifacts if the \ppp box in which that molecule is dissolved is large enough (\ie as all the distances from solute atoms to the box boundaries are larger than $R_\mathrm{cut,1}^{pol}$). Nevertheless we allow solute atoms to interact with \ppp periodic images, which can yield artifacts like over accumulation of charge like ions at the \ppp box boundaries as simulating a salt solution, for instance (that ion configuration  maximizes the polarization of \ppp's located close to the box boundaries). To prevent such spurious effects, we enforce all solute atoms to be confined in a sub volume of the \ppp box by means of the potential:

\begin{equation}
U_\mathrm{boundary}^{rep} =  k_\mathrm{boundary}^{rep} \sum_{\xi=x,y,z}  \left[ \max \left(2|\xi| - (L_\xi   - \delta L),0\right)  \right]^2,
\end{equation}
here $\xi$ are the atomic cartesian coordinates, $(L_x,L_y,L_z)$ are the simulation box dimensions, and $\delta L$ is set to a large enough distance (usually we set it to $R_\mathrm{cut,1}^{pol}$). For the present study we set the harmonic constant $k_\mathrm{boundary}^{rep}$ to 10.0 \kcal \AA$^{-2}$.

The starting structure of a solute dissolved in a \ppp box is built by setting the solute center of mass to the box center. Then we add \ppp's on a cubic grid (the distance among the grid nodes is 2.8 \AA). Lastly we remove all the \ppp's that lie at a distance $<$ 3.5 \AA $ $ from any solute non-hydrogen atom. 
 
\subsection{Long range electrostatics and the \pppl approach} \label{sec:long_range}

\begin{figure}
\includegraphics[scale=.8]{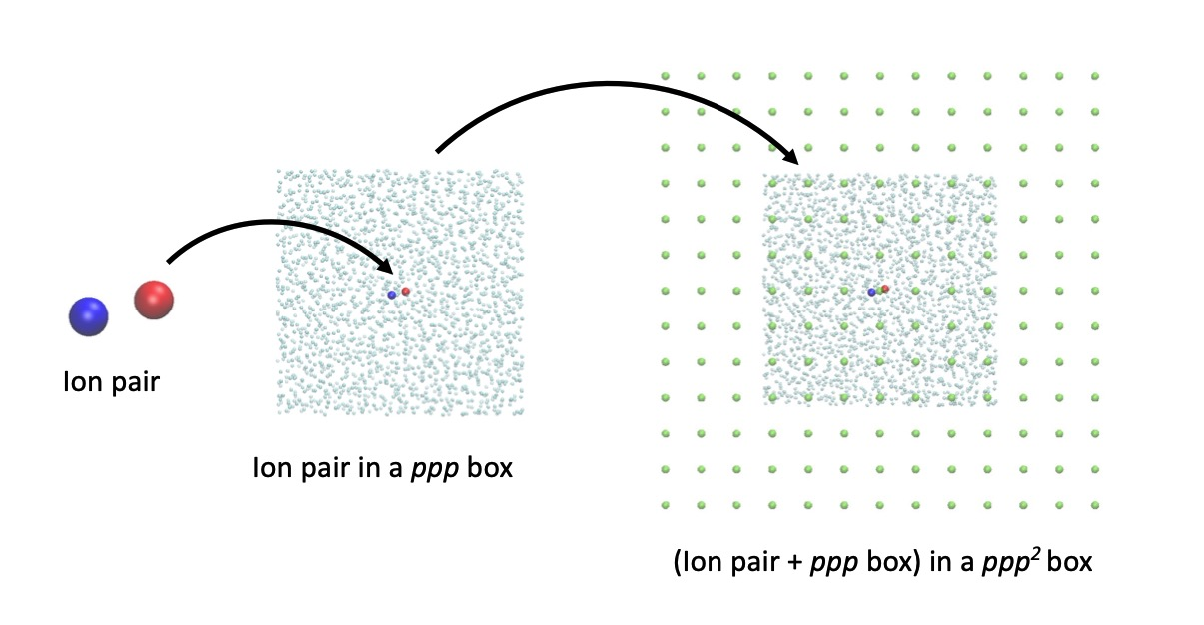}
 \caption{The multi-scale \pppl approach. Here an ion pair (blue and red spheres) is dissolved in a box of \ppp's (each \ppp models a single water molecule). Then the ion pair and the \ppp box is set at the center of a new box made of $ppp^2$ particules, each of them models a local spherical volume of the solvent water. The new ion pair + \ppp's + $ppp^2$'s systems may be then set in a larger $ppp^3$ box, up to reach a desired precision as computing long range solute/solvent electrostatic interactions.} 
 \label{fig:multiscale}
\end{figure}

We showed the effect of truncating the solute electric field for atom/\ppp distances larger than $R_\mathrm{cut,1}^{pol}$ to largely overestimate ion pair association in a \ppp fluid  as compared to real water \cite{masella13}. To remediate  that artifact we proposed the 'Russian doll' multi scale \pppl approach depicted in Figure \ref{fig:multiscale}. We add  $l \geq 2$ levels of larger and larger \ppp-like particles (denoted as \pppl,  $\ppp^1$ particles are the original \ppp's). Even if different choices may be done, we set here the \pppl radius to that of the original \ppp's scaled by an integer $n = 2^{l-1}$. A single \pppl particle thus models a cluster of $8^{l-1}$ original \ppp's. 

Particles of a given level $l$ (they are all of the same size) do not interact with the particles of a different level. Moreover they only electrostatically interact with a solute if at least one atom/\pppl distance $r$ obeys $R_{\mathrm{cut},l-1}^{pol} \leq r \leq R_{\mathrm{cut},l}^{pol}$. The solute/\pppl electrostatic energy term for $l>1$ corresponds to the linear version (for weak solute electric fields) of the polarization term of Equation~(\ref{eqn:ppp_polarization}) and dipole/dipole interactions among atoms and \pppl particles are omitted. We maintain the \pppl density along a simulation in the corresponding level box using periodic conditions. The numerical precision of a \pppl scheme can be estimated from Equation~(\ref{eqn:precision}) using the largest $R_{\mathrm{cut},l}^{pol}$ truncation distance.

For a given level $l >1$, \pppl's interact with each other according to the $\tilde{U}^{LJ}_\ppp$ potential detailed above, with rescaled $\sigma^*_l$ and $\epsilon^*_l$ parameters. For efficiency reason we truncate the \pppl interactions for inter particle distances larger than $5 \times l$ \AA.  In our original \pppl  approach, the $\sigma^*_l$'s were set to $\sigma^*_\ppp$ scaled by the ratio of the particles radius between levels $l$ and 1, whereas all the $\epsilon^*_l$'s were set to a low value to speed up the \pppl diffusion. In a recent study \cite{masella21}, we showed the latter assumptions to yield potential  drawbacks as simulating highly charged polyelectrolyte systems in presence of counter ions, \ie \pppl's over accumulate in between the polyelectrolyte and the counter ion cloud. To prevent such phenomena we arbitrarily enlarged the $\sigma^*_l$'s in the latter study.

In the present study we re assigned the parameters ($\sigma^*_l$,$\epsilon^*_l$) to reproduce the water density at ambient conditions for a neat \pppl fluid (taking into account a \pppl to correspond to a cluster of $8^{l-1}$ real water molecules). We set all the $\epsilon^*_l$'s to the value $\epsilon^*_\ppp$ of the original \ppp's, and we assigned the $\sigma^*_l$'s to reproduce the water density.  To meet the latter condition, the $\sigma^*_l$'s need to obey the basic relation $\sigma^*_l = \gamma \times 2^{l-1} \times \sigma^*_\ppp$, where $\gamma$ is a parameter set to 1.245615. 

Regarding the \pppl masses, we set them all to $2^3$ times the mass of an original \ppp (\ie the mass of a real water molecule), regardless of $l$. Such a large mass value slow down the \pppl diffusion and  that allows us to account for solute/\pppl long range  forces  as slow fluctuating interactions in the multiple time step algorithm to solve the Newtonian equations of motion detailed below.

For readability purpose, simulations performed using only first level \ppp particles are denoted below as performed in a neat \ppp fluid, whereas they are denoted as performed in a \pppl fluid as using higher level \pppl particles. 

\subsection {Intra solute interactions} \label{sec:ion}

In the present study we simulate explicit  ions \xx $ $ and \yy $ $ as polarizable mono atomic  centers. The monovalent ions are \na and \cl whose interactions among them and explicit  water molecules are modeled by means of the \allatom polarizable force field detailed in an earlier study \cite{vallet22}. Ions with higher Coulombic charges (Q $\geq$ 2 $e$) are modeled using the \na/\cl \allatom force field parameter set. However we reinforce ion/water short range repulsion to prevent unrealistic too short ion/water distances in MD simulations. Simulations of a hydrophobic polyelectrolyte polymer are performed using the polarizable \allatom force field detailed in our earlier study \cite{masella21}.
 
\subsection {MD simulation details} \label{sec:simulation-details} 

Preliminary (relaxation) MD simulations are performed only in a neat \ppp fluid and in the NPT ensemble, whereas production MD runs in \ppp and \pppl fluids are performed in the NVT ensemble. All simulations based on a \allatom force field are performed in the NPT enesemble. Temperature and pressure are monitored along NPT runs using the Nos\'e-Hoover barostat \cite{martyna96} (the barostat coupling constant is set to 2.5 ps), whereas temperature is monitored along NVT runs according to a Langevin thermostat approach \cite{cances07}. If not otherwise stated, induced dipole moments are solved iteratively until the mean difference in their values between two successive iterations is smaller than 10$^{-6}$ Debye and the maximum difference for a single dipole is smaller than $25\times10^{-6}$ Debye. The equations of motion are solved using a Multiple Time Steps, MTS, algorithm devoted to polarizable force field based on induced dipole moments \cite{masella06}. For NaCl simulations, two time steps are used: 2 and 6 fs for short range and long range electrostatic/polarization forces, respectively. To simulate the hydration of more complex entities (with intra molecular chemical bonds, like a polymer or the tetra methyl ammonium cation) we used three time steps, namely 0.25 fs for intra molecular stretching and bending forces, and 1.0/5.0 fs (small systems) or 2.0/6.0 fs (large systems) for short and long range electrostatic and dispersion forces.  1-4 dihedral forces are considered as short range forces and the cutoff distance to compute electrostatic/dispersion short range forces is set to 8 \AA, regardless of the solute. As discussed above, the forces corresponding to solute/\pppl interactions are systematically considered as long range polarization forces.

For \pppl simulations, the starting coordinates of solute atoms and \ppp's correspond to the final point of a preliminary  5 ns NPT simulation in a neat \ppp fluid.  All molecular modeling computations and simulations were performed with our own code POLARIS(MD) \cite{polaris}. 

 \subsection {Hydration free energy and PMF computations} \label{sec:TI_pmf}

Ion Gibbs hydration free energies $\Delta G_\mathrm{hyd}$ are computed in two steps using a 32 windows Thermodynamical Integration, TI, scheme \cite{frenkel02}. The first step consists in linearly downscaling to zero the ion charge and polarizability, and during the second step, the uncharged and non polarizable ion is then linearly transformed into a ghost entity fully decoupled from the solvent. To prevent numerical instabilities  during the second step, we add the quantity $1 - \lambda$ to all ion/\ppp distances  as computing solute/solvent interactions ($\lambda$ is the scaling parameter monitoring the progressive  ion/\ppp decoupling).  Each TI MD simulation is performed at the 2.5 ns scale and the ion/\ppp potential energy derivatives are computed each 250 fs once a starting relaxation phase of 0.5 ns is achieved. The two free energy components computed from the two main steps (whose sum yields $\Delta G_\mathrm{hyd}$) are denoted $\Delta G_\mathrm{pol}$ and $\Delta G_\mathrm{np}$, respectively.

The Potential of Mean Force, PMF, of an ion pair is computed using an umbrella sampling MD protocol. The degree of freedom $R$ restrained along MD simulations, using the harmonic potential $k_c \left(R - R_c\right)^2$, is the distance between the ion centers of mass. The target distances $R_c$ usually span from 2 to 17 \AA $ $ and they are regularly spaced by 0.5 \AA,. If not otherwise stated, the constant $k_c$ is set to 5 \kcal \AA$^{-2}$. The MD duration is set to 10 ns and the distances $R$ are sampled each 50 fs once a starting phase of 1 ns is achieved.  We  post process the set of sampled distances  to compute the PMF's according to the Umbrella Integration method \cite{kastner05}. The PMF's account for the entropic correction $2\mathrm{RT} \ln R$.

\section{Results}

In the following discussions, we denote as '\allatom', '\ppp' and '\pppl' data computed along \allatom simulations and simulations performed in a \ppp  or \pppl fluids, respectively. We denote as Coulombic potential the classical effective potential $qq' / 4\pi \epsilon r$ of two charges $q$ and $q'$ lying at a distance $r$ from each other and dissolved in a fluid whose dielectric constant is $\epsilon$.

\subsection{\ppp response to solute medium range electrostatics}

For each of the four ion pairs [\xx,\yy] defined in Section \ref{sec:ion}, we performed a 200 ns scale \allatom simulation in bulk water at ambient conditions. The distance between the ions \xx $ $ and \yy $ $ is harmonically restrained to a target distance of 14 \AA $ $ (the harmonic constant is set to 50 \kcal \AA$^{-2}$). We computed the mean water normalized density $\bar{\rho}_s$ and the mean dipole moment $\bar{\mu}_\mathbf{X}$ (projected on the ion pair axis and expressed in Debye per water molecule/\ppp) within a cylindrical volume centered at the ion pair center and whose length and radius are 2 and 0.5 \AA, respectively.  

We computed accordingly the quantities $\bar{\rho}_s$ and $\bar{\mu}_\mathbf{X}$ along 200 ns \ppp simulations performed using three dipole saturation $\mu_s$ values: 1.2, 2.0 and 12.0 Debye. The remaining ion/\ppp parameters are those of the model corresponding to the damping parameters $a_i = 0.3$ \AA$^{-3}$ detailed in the following Section \ref{sec:models}. As the ion/\ppp electric field damping vanishes beyond 5.0 \AA, the magnitudes of the quantities $\bar{\rho}_s$ and $\bar{\mu}_\mathbf{X}$ do not depend on the damping parameters $a_i$'s. 

The $\bar{\rho}_s$ and $\bar{\mu}_\mathbf{X}$ values for \allatom and \ppp simulations are summarized in Table \ref{tab:mean_data}. The magnitude of $\mu_s$ has a weak effect on the densities $\bar{\rho}_s$  along \ppp simulations. Those densities depend mostly on the ion charge $Q$: they increase from 1.03 ($Q=1$) to 1.20  ($Q=4$) $\pm$ 0.01, regardless of $\mu_s$. Along \allatom simulations, the ion charge has a weaker but opposite effect on  $\bar{\rho}_s$ : it decreases from 0.97 to 0.90 $\pm$ 0.01 as $Q$ increases. 

\begin{figure}
\includegraphics[scale=.7]{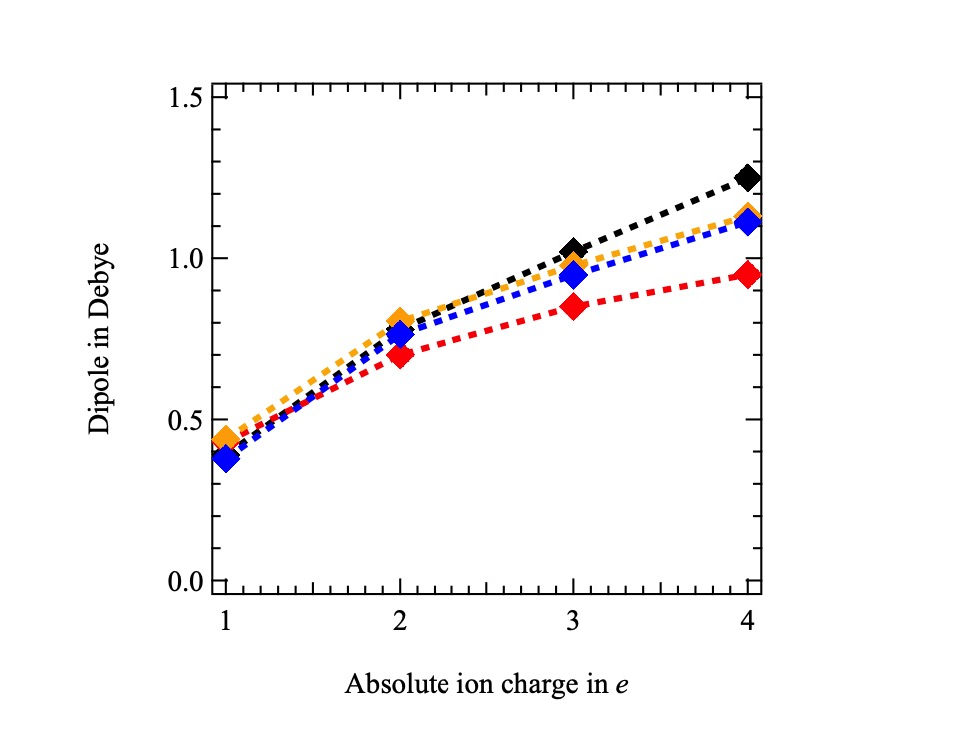}
 \caption{Water and \ppp mean dipole projections $\bar{\mu}_\mathbf{X}$ and mean dipole projections scaled by the solvent mean density $\bar{\rho}_s$ as a function of the ion absolute charge $Q$. \ppp data correspond to $\mu_s = 1.2$ Debye. Black (red) and blue (orange) symbols: water (\ppp) dipole projections and density scaled data, respectively.} 
 \label{fig:dipole_ppp}
\end{figure}

Contrary to the densities $\bar{\rho}_s$, $\mu_s$ has a much more significant effect on the \ppp dipole projections $\bar{\mu}_\mathbf{X}$: they increase by a factor ranging from 2  ($\mu_s = 1.2$ Debye) up to  4 ($\mu_s = 12.0$ Debye) as $Q$ increases from 1 to 4. We also note a good agreement between the  \ppp  $\bar{\mu}_\mathbf{X}$ data corresponding to $\mu_s = 2.0$ Debye and \allatom values. However as the \ppp polarizability depends explicitly on the solvent density, we need to compare the $\bar{\mu}_\mathbf{X}$ values scaled by the the local solvent densities to draw sound conclusions. As shown by the plots of Figure \ref{fig:dipole_ppp}, the \ppp values $\bar{\mu}_\mathbf{X} \times \bar{\rho}_s$ for $\mu_s = 1.2$ Debye (a value that we considered in our earlier studies \cite{masella21,masella23}) nicely match the corresponding \allatom data.

The parameter $\mu_s$ was originally introduced to prevent potential over polarization effects at short range from a solute. In our \ppp approach, that parameter allows us to readily compensate \ppp over concentration effects at medium range from heavily charged solutes. Note that our estimate of  the isothermal compressibility of a neat \ppp fluid, about 16 $\pm 1$ 10$^{-6}$ atm$^{-1}$ for pressures ranging from 1 to 1000 atm and T = 300 K, is three times smaller than for neat water. 

 \begin{table}[htp]
\caption{Water and \ppp densities $\bar{\rho}_s$ (normalized to the bulk water density) and dipole projections $\bar{\mu}_\mathbf{X}$ on the ion axis (in Debye per water molecule/\ppp) at the centers of a dissociated [\xx,\yy] ion pair.  The \allatom projections $\bar{\mu}_\mathbf{X}^0$ is computed from water static dipoles. The saturation parameter $\mu_s$ is expressed in Debye. The standard deviations are about 0.15 (\allatom) and 0.22 (\ppp) for density data, and about 0.05 (\allatom) and 0.005 (\ppp) Debye per molecule/\ppp  for dipoles. The parameters of the \ppp energy terms $\tilde{U}^{LJ}_\ppp$ and $U_{dens}$ are $\sigma^*_\ppp = 2.85$ \AA, $\epsilon^*_\ppp = 0.53$ \kcal, $(\nu_1,\bar{n}_1) = (0.15,4.2)$ and $(\nu_2,\bar{n}_2) = (0.0475,15.0)$ (those parameters are expressed in \kcal and \ppp number, respectively).} 
\label{tab:mean_data}
\begin{center}
\begin{tabular}{ccccccccc}
\hline
\hline 
    Charge $Q$ & \multicolumn{2}{c}{\allatom} &  \multicolumn{2}{c}\ppp($\mu_s=1.2$)  &  \multicolumn{2}{c}\ppp($\mu_s=2.0$) & \multicolumn{2}{c}{\ppp($\mu_s=12.0$)} \\
              & $\bar{\rho}_s$ & $\bar{\mu}_\mathbf{X}$ ($\bar{\mu}_\mathbf{X}^0$) & $\bar{\rho}_s$ & $\bar{\mu}_\mathbf{X}$ & $\bar{\rho}_s$ & $\bar{\mu}_\mathbf{X}$ & $\bar{\rho}_s$ & $\bar{\mu}_\mathbf{X}$  \\
\hline 
\hline 

1 & 0.97  &   0.39 (0.28) & 1.02 &  0.43 & 1.03 & 0.45 & 1.04 & 0.46 \\
2 & 0.98 & 0.78 (0.55) & 1.15 &  0.71 & 1.15 & 0.83 & 1.17 & 0.93 \\
3 & 0.93 & 1.02 (0.71) & 1.15 &  0.85  & 1.16 & 1.11 & 1.16 & 1.39 \\
4 & 0.90 & 1.25 (0.88) & 1.19 &   0.95 & 1.20 &  1.30 & 1.21 & 1.85   \\ 

  \hline
\hline
\end{tabular}
\end{center}
\label{default}
\end{table}

 \clearpage

\subsection{\pppl domain boundary effects on long range electrostatics} \label{sec:boundary_effects}

In the multi-scale \pppl approach, solute/solvent long range electrostatics is taken into account by a sum of solvent domain effects. Each domain corresponds to set of particles of increasing size, particles that can over accumulate at the domain boundaries to strengthen solute/solvent interactions as discussed in an earlier study \cite{masella13}.

To investigate potential drawbacks arising from particle over accumulation at \pppl domain boundaries, we performed two series of simulations of the cations \na and \xb, as well as of a single [\na,\cl] ion pair as dissolved alone in a two level \pppl fluid. Each \pppl box comprises about 2 000 particles and the largest cutoff distance $R_{\mathrm{cut},2}^{pol}$ is set to  31 \AA. The two simulation series correspond to $R_{\mathrm{cut},1}^{pol}$ set to 12 and 15 \AA, respectively. Regarding the ion pair and for each $R_{\mathrm{cut},1}^{pol}$ value, we performed three simulations along which we harmonically restrain the [\na,\cl] distance to 3, 5 and 8 \AA, respectively. From the particle normalized radial distribution functions $g^l_\ppp(r)$ computed from the cationic centers along those simulations, we compute the functions $\Delta \rho^l(r) = g^l_\ppp(r) - 1$ (that we express in \%). Below we denote as boundary domain the solvent shell domain corresponding to $R_{\mathrm{cut},1}^{pol} \pm 0.5 $ \AA. The ion electric field is progressively zeroed (original \ppp's) or increased from zero to its unaltered value (\pppa's) within the latter domain.

\begin{figure}
\includegraphics[scale=.65]{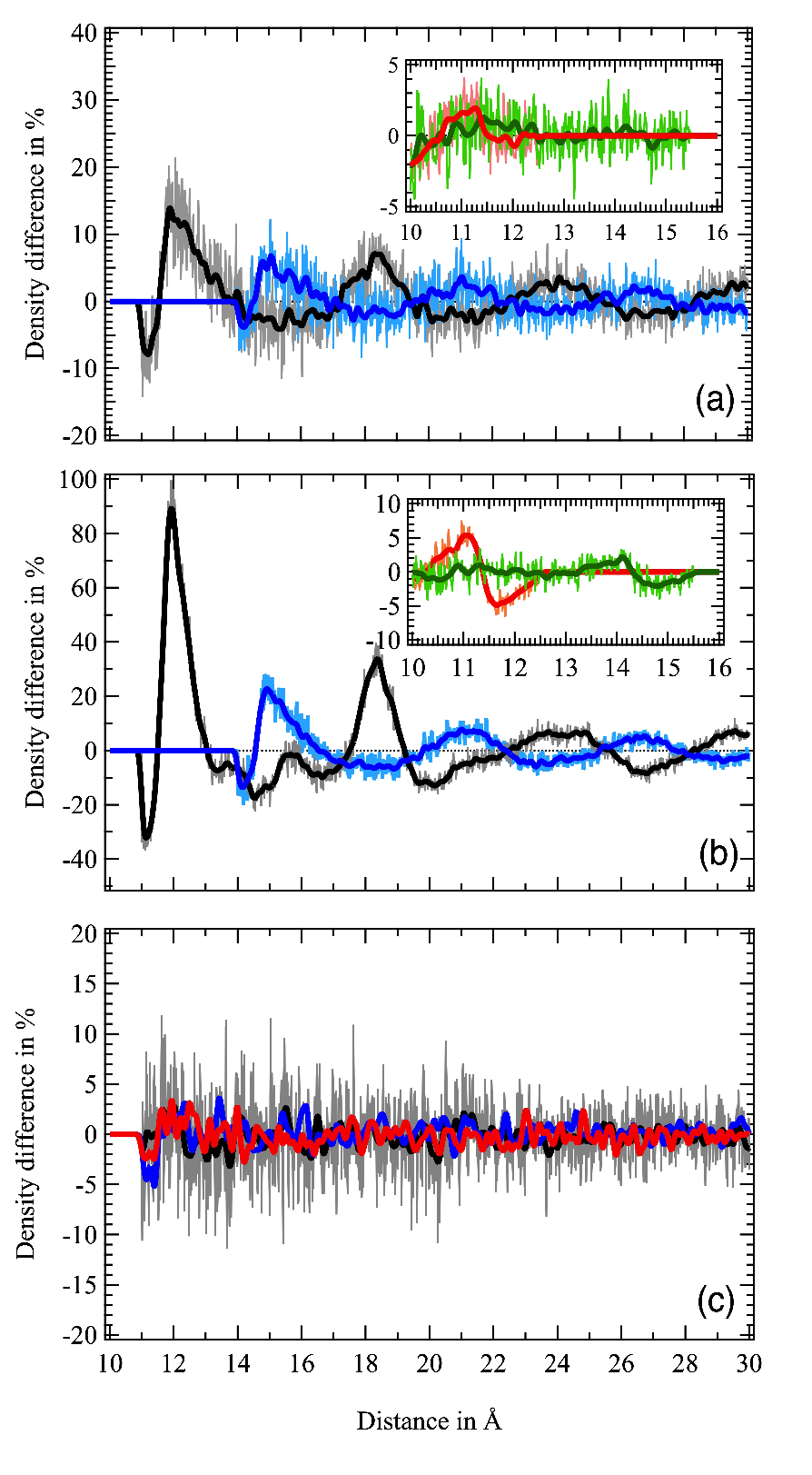}
 \caption{Differences  $\Delta \rho^l(r)$ in the local \pppa particle density. (a): from the cation \na dissolved alone in a 2 level \pppl fluid; (b): from the cationic center \xb. For (a) and (b) the black and blue lines correspond to $R_{\mathrm{cut},1}^{pol} = $ 12 and 15 \AA, respectively. The insets show the density differences for original \ppp particles (red and green lines: data corresponding to $R_{\mathrm{cut},1}^{pol} = $ 12 and 15 \AA). (c): difference in the local \pppa particle density from \na for a [\na,\cl] pair dissolved in a 2 level \pppl fluid. Black, blue  and red: the [\na,\cl] distance is harmonically restrained to 3, 5 and 8 \AA $ $ along the simulations. Thin and bold lines: raw data and the corresponding smoothed data from a binomial scheme. The density differences are here arbitrarily zeroed for particle/cation distances larger (\ppp's) and smaller (\pppa's) than  $R_{\mathrm{cut},1}^{pol}$. The mean density $\bar{\rho}_s^l$ for pure \ppp and \pppa fluids are 3.335 and 0.419 10$^{-2}$ particles per \AA$^3$, respectively.} 
 \label{fig:density_boundary}
\end{figure}

We plot the functions $\Delta \rho^l(r)$ in Figure \ref{fig:density_boundary}. For single cations, those plots show \ppp and \pppa particles to over accumulate within the boundary domain. The magnitude of the particle over accumulation increases as the cation charge increases and/or as the $R_{\mathrm{cut},1}^{pol}$ distance decreases. Within the boundary domain,  the \ppp over accumulation is overall weak, it never exceeds 5 \%, whereas the \pppa over accumulation is much more accented. $\Delta \rho^2(r)$ is as large as +90\% (\xb) and +15\% (\na) using  $R_{\mathrm{cut},1}^{pol}=12$ \AA, and it amounts to 20\% (\xb) and +6\% (\na) for $R_{\mathrm{cut},1}^{pol}=15$ \AA. Moreover  we note the $\Delta \rho^2(r)$'s for single cations to oscillate for distances above the boundary domain, showing the cation effects on the \pppl fluid structure to extend at long range from them. 

For [\na,\cl] pairs and regardless of the cation/anion target distance, the magnitude of the differences $\Delta \rho^2(r)$ is much weaker compared to single cations. It amounts at most to +3 \%  within the boundary domain and then it converges rapidly to zero on average. That suggests boundary artifacts to be weak as modeling common salt solutions using the \pppl approach.

As \ppp over accumulation is taken into account as assigning solute/solvent parameters (see below), its effect on hydration energies is removed. Regarding \pppa's we estimate the energetic error arising from their over accumulation at the boundary domain as the difference $\Delta U_{ps}^2$ in the cation/\pppa electrostatic energy computed from the functions  $\Delta \rho^2(r)$. $\Delta U_{ps}^2$ is computed from a relation similar to Equation (\ref{eqn:precision}):

\begin{equation}
\Delta U_{ps}^2 = - \frac{\alpha_{s,2} \rho_{s,2}}{ \epsilon_0} \int_{R_\mathrm{cut,1}^{pol}}^{R_\mathrm{cut,2}^{pol}}  \Delta \rho^2(r)\frac{f(r)^2}{r^2}  dr, 
\end{equation}
here, $\alpha_{s,2}$ and $\bar{\rho}_{s,2}$ are the \pppa polarizability and mean density, respectively. $f$ is the truncation function $G$ at the \ppp/\pppa boundary domain. The expected electrostatic cation/\pppa energies $U_{ps}^{2}$ (computed from Equation (\ref{eqn:precision}) by setting the lower and upper bounds to $R_{cut,1}^{pol}$ and $R_{cut,2}^{pol}$ and by accounting for the smoothing function $G$) for a single cation \na are 8.3 and  5.5 \kcal  for $R_{\mathrm{cut},1}^{pol} = 12$ and 15 \AA, respectively. The magnitude of the $U_{ps}^{2}$ values for \xb are four times larger than the latter ones. 

The \pppa over accumulation at the boundary domain yields $\Delta U_\mathrm{pol}^2$ to amount from 4\% (\na) up to 14\% (\xb) of the corresponding $U_{ps}^{2}$ data for $R_{\mathrm{cut},1}^{pol} =  12$ \AA, and by about 1.5 and 4\% for the latter cations for $R_{\mathrm{cut},1}^{pol} =  15$ \AA. From the \na/\pppa radial distribution functions corresponding to our [\na,\cl] pair simulations,  $\Delta U_{ps}^2$ amounts at most to 0.05\% of the expected $U_{ps}^{2}$ value. Domain boundary effects thus yield strong drawbacks only for highly charged solutes as modeled at infinite dilution conditions and in absence of counter ions. 

Boundary domain artifacts on ion/\pppa electrostatic energies can be minimized by considering a large enough  $R_{\mathrm{cut},1}^{pol}$ value. However an alternative and computationnaly more efficient route to minimize boundary domain effects on hydration energy is to alter the solute/solvent electrostatic energy $u_{ps}^l$ for each \pppl particle according to :

\begin{equation}
u_{ps}^l = -\alpha_{s,l} E^2_s \times ( 1 - \eta E^2_s),
\end{equation}
here $E^2_s$ is the square of the electric field magnitude generated by a charged solute on a \pppl particle, regardless of the level $l \geq 2$. By setting the parameter $\eta$ to 670.0 \AA$^4$ $e^{-2}$, the error $\Delta U_\mathrm{pol}^2$ for the single cations \na and \xb is then about 2\% and 1\% of the $U_\mathrm{pol}^2$ values as using $R_{\mathrm{cut},1}^{pol} =  12$ and 15 \AA, respectively. The above correction scheme has a negligible effect as modeling an ion pair like [\na,\cl]. We systematically consider that approach and the above value of the parameter $\eta$  in all the simulations discussed below.

\begin{figure}
\includegraphics[scale=.8]{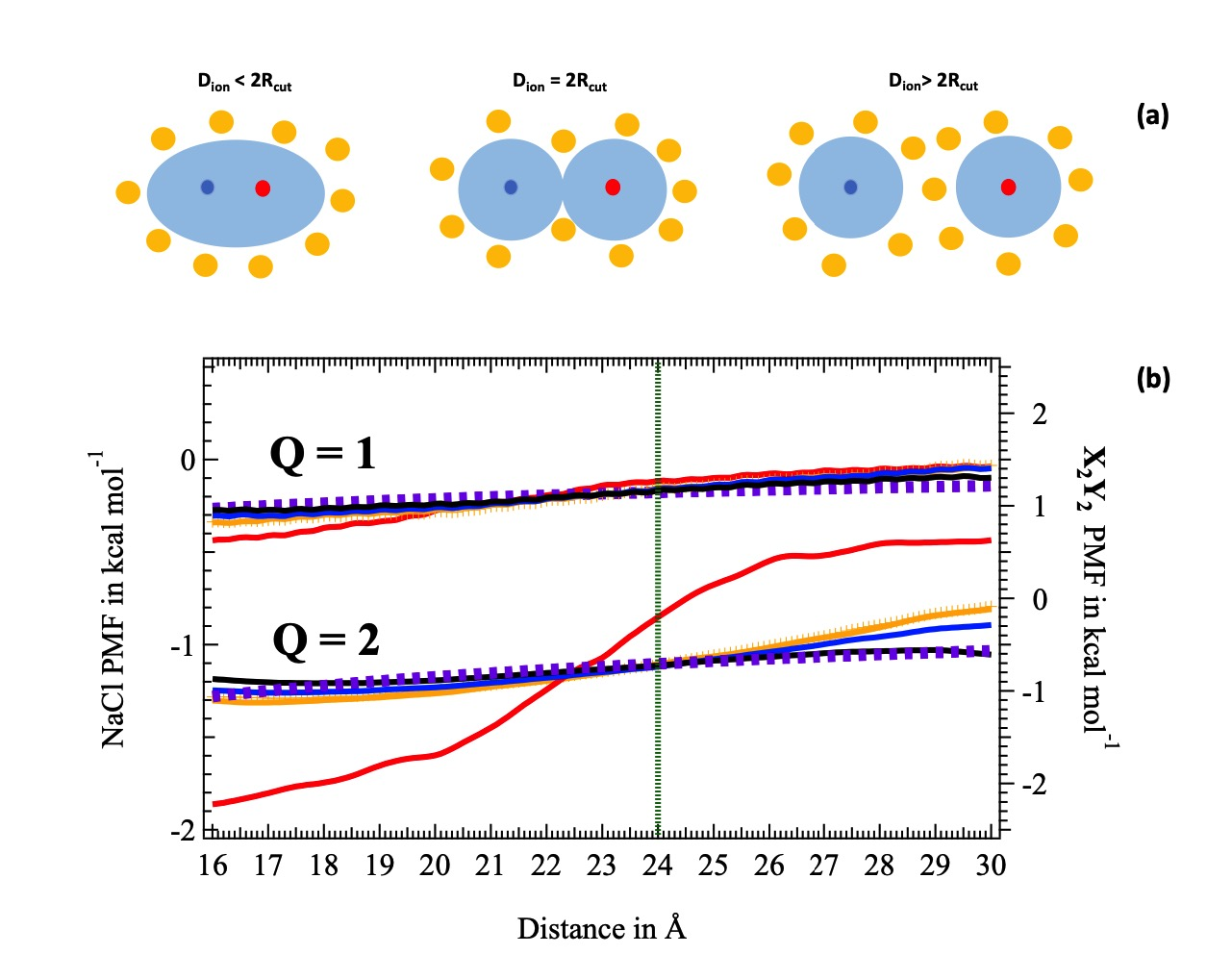}
 \caption{Ion pair PMF's in \pppl fluids as a function of the truncation distance $R_{cut,1}^{pol}$. (a) schematic representation of the solvent elements explicitly taken into account \emph{wrt}   $R_{cut,1}^{pol}$. Blue: \ppp particles represented as a continuum medium. Yellow spheres: \pppa particles. D$_\mathrm{ion}$ is the ion separation distance and $\mathrm{R}_\mathrm{cut} = R_{cut,1}^{pol}$. (b) PMF's of the [\na,\cl] (right axis) and [\xb,\yb] (left axis) ion pairs as dissolved in \pppl fluids. Violet dashed lines: expected Coulombic potentials for an opposite ion pair (the absolute charge of both ions is Q). Red, yellow (or yellow cross symbols), blue and black lines: PMF's computed by setting  $R_{cut,1}^{pol}$ to 12, 15, 18 and 21 \AA, respectively. All PMF's are computed in a four level \pppl fluid, at the exception of those shown by small cross symbols, which correspond to a six level \pppl fluid.}
 \label{fig:pmf_long_range}
\end{figure}
 
We discuss more in details the role of accounting for more and more \pppl levels on ion pair PMF's in the following section. Anticipating our findings, the expected Coulombic potential for an ion pair [\na,\cl] dissolved in neat water for intermediate 8 to 16 \AA $ $ ion separation distances is well reproduced  by taking in account at least three \pppl levels. Assuming that result, we investigated possible artifacts arising from the presence of domain boundaries on the long range tail of the PMF's of the ion pairs [\na,\cl] and [\xb,\yb], in particular at ion separation distances close to twice the value of $R_{\mathrm{cut},1}^{pol}$. For such ion separation distances, the \pppa's located in between the ions start to explicitly interact with them, see Figure \ref{fig:pmf_long_range}(a). We performed a new series of 10 ns scale simulations of the ion pairs as dissolved in 7 000 \pppl's boxes and by setting $R_\mathrm{cut,1}^{pol}$ to 12, 15, 18 and 21 \AA, respectively. Long range solvent effects are taken into account by considering  4 and 6 levels \pppl schemes. These new PMF's are compared to the expected Coulombic potentials for the two ion pairs in Figure \ref{fig:pmf_long_range}(b). In that Figure, the PMF's are switched to minimize their numerical difference with the Coulombic potential for ion separation distances spanning from 16 to 30 \AA.

All the new [\na,\cl] PMF's match the expected Coulombic potential on average, within less than 0.1 \kcal (regardless of $R_\mathrm{cut,1}^{pol}$) and even 0.03 \kcal (for $R_\mathrm{cut,1}^{pol} \geq 15$ \AA) on the 16-30 \AA {} distance range. However  $R_\mathrm{cut,1}^{pol}$ has a stronger effect  on the [\xb,\yb] PMF's. For instance the PMF computed for $R_\mathrm{cut,1}^{pol} = 12$ \AA $ $ (and by taking into account 4 \pppl levels) obeys two regimes in the distance domain 16-30 \AA, with a step transition between them at about 24 \AA $ $, \textit{i.e.} at twice the $R_\mathrm{cut,1}^{pol}$ value, a distance at which higher level \pppa's located in between both ions start to interact with them. Increasing the cutoff distance $R_\mathrm{cut,1}^{pol}$ to values $>$ 12 \AA $ $ yield a better agreement with the expected Coulombic potential (within less 0.05 \kcal for $R_\mathrm{cut,1}^{pol} = 21$ \AA, for instance).

Domain boundary effects on complex ionic solutions are expected to play a marginal role because of our shell truncation scheme. The presence of a 'real' molecular/ionic entities at the vicinity of an ion pair center would usually prevent to account for interactions between the ion pair and the high level \pppl particles located in between those ions. As modeling the hydration of an assembly made of two independent molecular systems (and whose largest absolute total charges are larger than 1 $e$), we just need to set $R_\mathrm{cut,1}^{pol}$ to a value larger than half the smallest inter atomic distance between the two systems to minimize the weight of the potential domain boundary artifacts.

\subsection{Short range electric field damping and  \ppp models} \label{sec:models}

To model the hydration of ions \na and \cl, we built a series of \ppp models for which we systematically set $\mu_s$ to 1.2 Debye. Those models differ by the value of the ion/\ppp damping parameter $a$ discussed in Section \ref{sec:cg}, which is taken identical for both ions. Here we consider the set of $a$ values 0.03, 0.05, 0.08, 0.10, 0.15, 0.20 and 0.30 \AA$^{-3}$.   For each model, we assign the parameters of the ion/\ppp energy term $\tilde{U}^{LJ}$ to reproduce the ion/water oxygen distances in ion first hydration shells from our earlier \allatom simulations \cite{vallet22} (\ie 2.4 and 3.4 \AA $ $  for \na and \cl, respectively) and to reproduce the ion experimental hydration Gibbs free energies $\Delta G^0_\mathrm{hyd}$ of \citet{kelly06} within 0.1 \kcal. Those target energies are -103.8 (\na) and -74.5 (\cl) \kcal. The \ppp energies $\Delta G_\mathrm{hyd}$ are computed according to the TI protocol detailed in Section \ref{sec:TI_pmf}. The simulated systems are single ions dissolved alone in a cubic box comprising 1 000 \ppp's. The hydration energies account for the long range polarization correction of Equation (\ref{eqn:precision}), which amounts to -13.5 \kcal. 

\begin{figure}
\includegraphics[scale=.7]{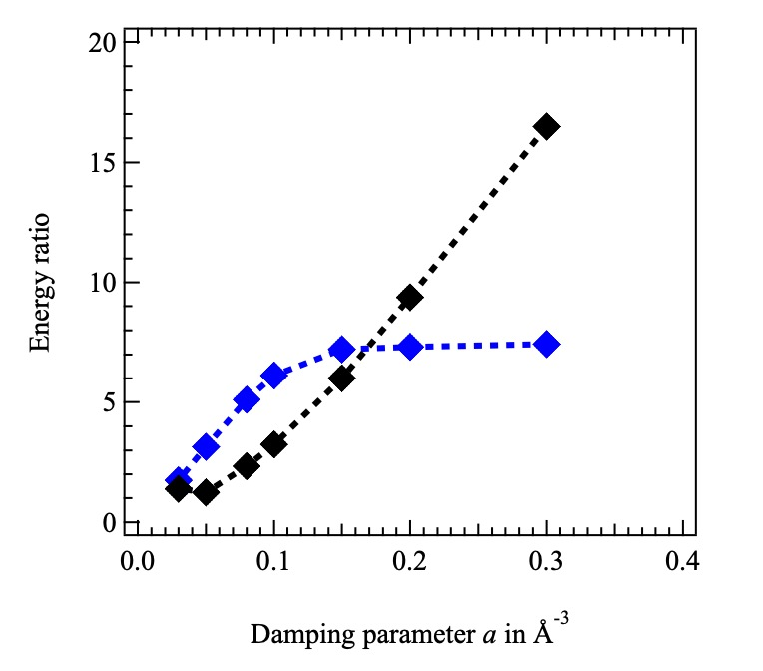}
 \caption{Ratio of the ion hydration free energy components $\Delta G_\mathrm{pol}$ and $\Delta G_\mathrm{np}$ as a function of the damping parameter $a$. Black and blue symbols: \na and \cl data, respectively.} 
 \label{fig:hydration_energy}
\end{figure}
         
The two components $\Delta G_\mathrm{pol}$ and $\Delta G_\mathrm{np}$ of the hydration energies $\Delta G_\mathrm{hyd}$
(see Section \ref{sec:TI_pmf}) for all \ppp models are negative and their ratios are plotted as a function of the damping parameter $a$ in Figure \ref{fig:hydration_energy}. Because of the shortest ion/\ppp distances in \na first hydration shell, the ratio for that cation varies on a wider range of values (from 1.5 to 16.5) than for \cl (from 1.5 to 7.5). However for the weakest damping parameter $a = 0.03$ \AA$^{-3}$ (a value that corresponds to strong damping effects), the energy ratios are close for both ions, about 1.5.

All those models predict ion first coordination shells $n_c$ to comprise about 5-6 (\na) and 15-16 (\cl) \ppp's. For \na that is in line with experimental and \allatom data \cite{othaki93,othaki01,gomez22}, whereas our \cl estimate is about twice as large as all available data. Regarding the entropic contribution to the hydration free energies $\Delta G_\mathrm{hyd}$, it amounts to 45 cal mol$^{-1}$ K$^{-1}$, regardless of the ion, a value that is from twice to four times larger compared to available experimental-based data \cite{tissandier98,schmid00}. The inability of coarse grained approaches to usually predict well balanced enthalpic and entropic contributions to the hydration free energies have already been discussed, in particular by \citet{noid23}. As we proposed in an earlier study  \cite{masella13}, we can reduce the $n_c$ value for \cl using \emph{ad hoc} three body potentials. However, even by using such sophisticated potentials, we have not been able up to now to better reproduce the entropic contribution to the ion hydration free energies. 

\begin{figure}
\includegraphics[scale=.75]{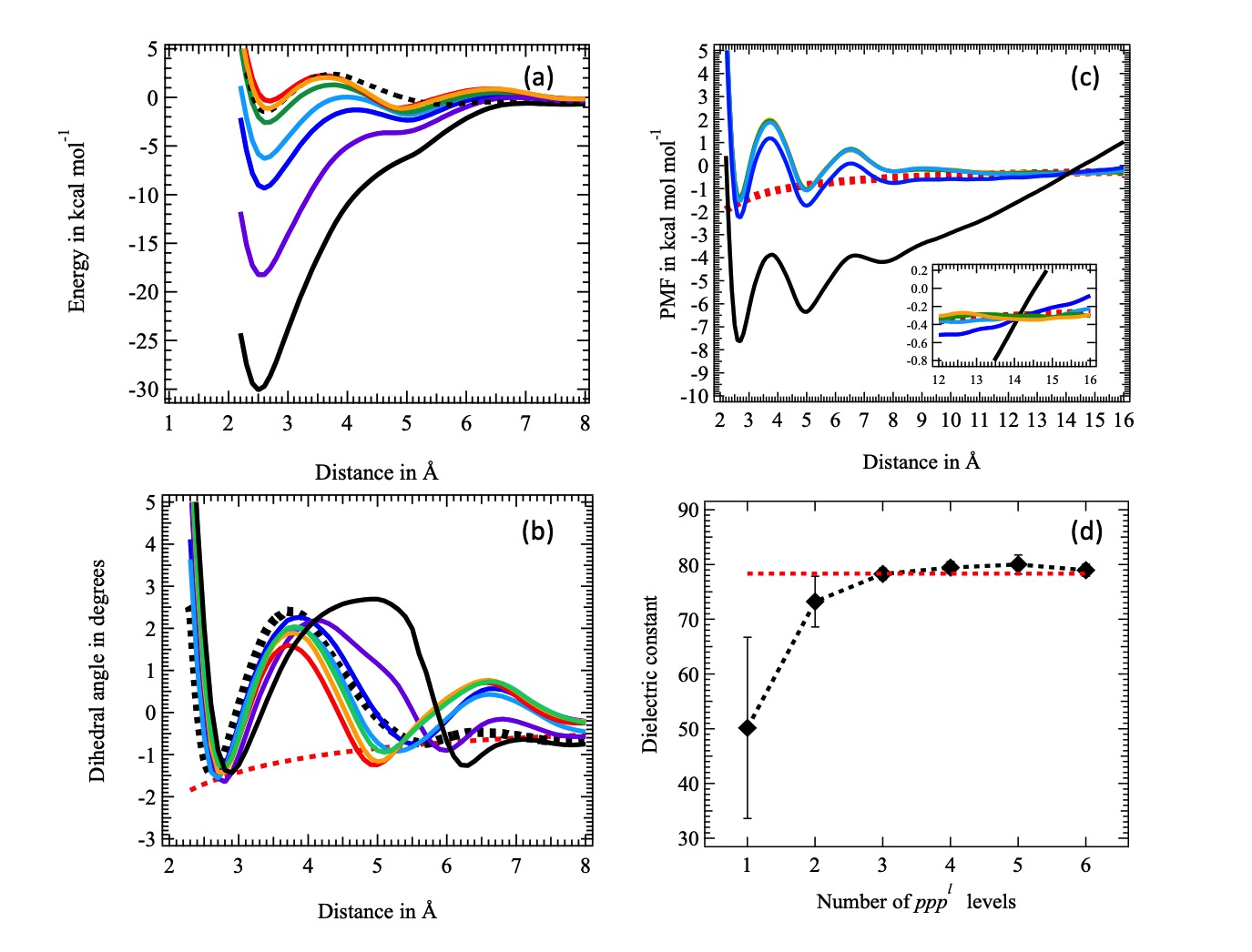}
 \caption{PMF's of the [\na,\cl] ion pair in \pppl fluids. All the PMF's are shifted to best reproduce the expected Coulombic potential of the ion pair in neat water for inter ionic distances that span between 12 to 16 {\AA}. From black to red bold lines: PMF's of \ppp models corresponding to a damping parameter $a$ that increases from 0.03 to 0.3 \AA$^{-3}$ (see text). (a) and (b): original \pppl PMF's and new PMF's computed using the correction energy term $\delta U^a$. Black dashed line:  [\na,\cl] PMF in neat water from \emph{all atom} simulations \cite{vallet22}. Red dashed line: Coulombic potential for an opposite ion pair in neat water (the water dielectric constant is here 78.35). (c): PMF's for a damping parameter $a = 0.15$ \AA$^{-3}$ \emph{wrt} the number of \pppl levels taken into account (black to orange: from one to four \pppl levels, red dashed line: Coulombic potential).  The inset shows the \pppl PMF's for inter ionic distances that span between 12 and 16 \AA. (d): convergence of the \pppl apparent dielectric constant $\tilde{\epsilon}_0^l$ as a function of the number of \pppl levels taken into account (red horizontal dashed line: the dielectric constant of neat water at ambient conditions, \ie 78.35). } 
 \label{fig:pmf}
\end{figure}

We computed the PMF's of an ion pair [\na,\cl] in a \pppl fluid from the 7 above models. The simulated systems correspond to a single ion pair dissolved in a 2 000 \ppp's box. Long range ion/solvent effects are explicitly accounted for using a four level \pppl approach, which allows us to account for ion/solvent interactions up to 150 \AA $ $ from the ions. Simulations are performed at the 10 ns scale. The resulting PMF's are compared to our \allatom  [\na,\cl] PMF \cite{vallet22} in Figure \ref{fig:pmf}(a). Whereas we are able to well reproduce the \allatom PMF when using large damping parameter values ($a \geq 0.2$ \AA$^{-3}$), the use of weaker and weaker damping parameter values yields to more and more over estimate the depth of the first PMF minimum corresponding to an associated ion pair. 

As shown by \allatom simulations \cite{debiec14}, the magnitude of the ion association process in the aqueous phase is tied to the depth of that first PMF minimum. In particular the deeper is that first minimum the stronger is the expected percentage of ion associated in bulk water. To best reproduce the [\na,\cl] \allatom PMF in water as using a \ppp model corresponding to a weak damping parameter $a$, we introduce a many-body correction potential $\delta U^a$ whose aim is to alter the [\na,\cl] PMF first minimum:

\begin{equation}
\delta U^a = \sum_i \left( \sum_{j\neq i} \frac{\xi_{ij}}{r_{ij}^2} \right) \times \left(  \sum_k G(r_{ik},R_{a},\delta R_a) \right).
\end{equation}
Here, $i$ and $j$ are two distinct ions and the $k$'s are the \ppp's. The sum on $k$ in the above relation is a measure of the local density of the \ppp's lying at short range from ion $i$. The function $G$ is defined in Equation (\ref{eqn:cutoff}). We adjust the parameters $\xi_{ij}$, $R_a$ and $\delta R_a$ to best reproduce the \allatom [\na,\cl] PMF in bulk water for each \ppp model. All the resulting new \ppp PMF's overall match the \allatom one, see Figure \ref{fig:pmf}(b). However as using $\delta U^a$, the second minimum of the new \ppp PMF's  is shifted to larger ion/ion distances (up to 6 {\AA}), and the height of the energy barrier interconnecting the PMF first and second  minimum is increased (up to reach 2.8 \kcal). Note the position of the second minimum in our \allatom PMF \cite{vallet22} is 5.5 \AA $ $ and the energy barrier height is 2.4 \kcal.

We compare in Figure \ref{fig:pmf}(c) the PMF's of a [\na,\cl] pair dissolved in \pppl fluids to its expected Coulombic potential for ion separation distances that span up to 16 \AA {} and as using the \ppp model corresponding to the  damping parameter $a = 0.15$  \AA$^{-3}$. Those PMF's are switched to minimize their numerical difference with the Coulombic potential for ion separation distances that span from 12 to 16 \AA. Note the PMF's computed from all the \ppp models to be indistinguishable within the latter ion separation distance range.  Moreover, within that distance range,  the ions only interact with a single shell of \ppp's and with higher level \pppl's that all lie outside that \ppp first shell. The new PMF's converge towards the Coulombic potential as soon as a 3 levels \pppl approach. By best fitting the new PMF's to a Coulombic-like potential $-1 / (4\pi \tilde{\epsilon}_0^l r_\mathrm{ion})$ ($r_\mathrm{ion}$ is the ion separation distance), we estimated the apparent dielectric constant $\tilde{\epsilon}_0^l$ of the \pppl fluids at intermediate ion separation distances.  $\tilde{\epsilon}_0^l$ is converged to the expected neat water value (78.35) as soon as 3 levels, within 1.5 \% (that corresponds to uncertainty of our fitted values for $l\geq3$).

\subsection{NaCl association in a \pppl fluid}

We investigate 0.2, 0.6 and 1.0M salty NaCl aqueous solutions using the different \ppp models detailed above. The simulated systems correspond to 200, 600 and 1 000 [\na,\cl] pairs dissolved in 100k \ppp cubic boxes, respectively. As the ions are confined in sub volumes of the \ppp boxes (see Section \ref{sec:cg}), the salt concentration is estimated from those sub volumes. To account for long range electrostatics, we consider a four level \pppl scheme.  For comparison purpose with our earlier \allatom data, the dipole moments are iteratively solved until the mean dipole moment difference between two successive iterations is smaller than 10$^{-6}$ Debye and the two time steps for solving the equations of motion (to handle short and long range interactions, respectively) are set to 2 and 6 fs. The trajectories are sampled each 12 ps once a starting relaxation phase of 5 ns is achieved. For readability purpose we discuss below energetic results using the unit \mcal, which corresponds to 10$^3$ \kcal.

For each salt concentration, we performed a 200 ns scale NVT simulation per \ppp model. We investigate ion clustering as in our study dealing about NaCl salty aqueous droplets \cite{vallet22}. At a time $t_0$ along a simulation, an ion cluster is a set of \na {} and \cl {} ions that are all located at distance shorter than $d_\mathrm{ref}$ from at least one another cluster ion. The cluster survives  until one of its ions leaves it or until a new ion is added to it. We set $d_\mathrm{ref}$ to 4.5 \AA, see below. For each cluster size $k$ we estimate the mean cluster survival time $ t_k^*$ from the correlation functions of the cluster survival probability at a time $t$ from $t_0$ \cite{vallet22}.  \pppl and \allatom clustering data are summarized in Table \ref{tab:clusters}.

\begin{figure}
\includegraphics[scale=.75]{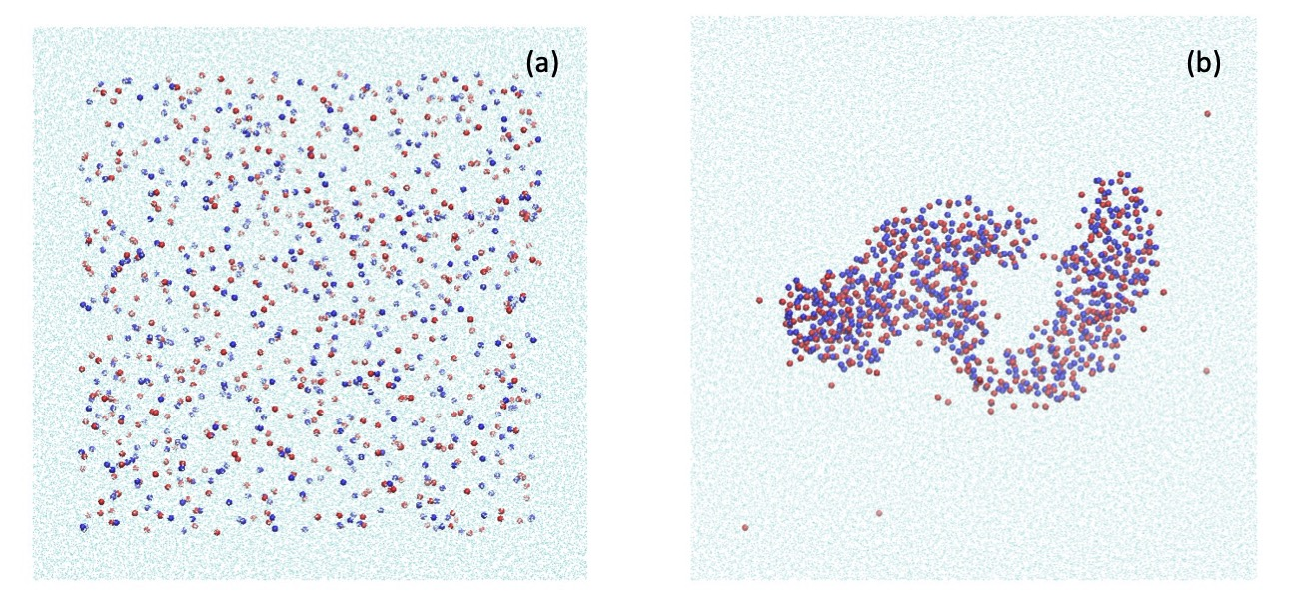}
 \caption{Two snapshots after  200 ns of simulations in a \pppl fluid of a 1.0M NaCl salt solution (see text for details). (a) and (b): snapshots from a damping parameter $a$ set to 0.05 (left) and 0.30 (right) \AA$^{-3}$. \ppp particles and ions \na and \cl are shown by light blue, dark blue and red spheres, respectively.} 
 \label{fig:snapshots}
\end{figure}

Regardless of the salt concentration, the ion/\ppp polarization damping parameter $a$ has a strong effect on the \na/\cl association in a \pppl fluid. Large values of $a$ ($\geq$ 0.20 \AA$^{-3}$) yield the formation of large clusters comprising most of the ions, whereas weak value of $a$ ($\leq$ 0.08 \AA$^{-3}$) yield the ions to be mostly free in solutions (see also the simulation snapshots shown in Figure \ref{fig:snapshots}). In the latter case from 80\% (0.6M and 1.0M salt solutions)  and 95\% (0.2M solutions)  of the ions are dissociated in a \pppl fluid. That is fully in line with  earlier \allatom simulations of salty aqueous bulk solutions and aqueous droplets performed based on polarizable force fields \cite{soniat16,vallet22}. 

 The order of magnitude of the cluster mean survival times $t_k^*$ for cluster size ranging from 2 to 4 (at the 0.1 ns scale, see Table \ref{tab:clusters}) are in line with available \allatom estimates, at the remarkable exception of 0.2M and 0.6M salt solutions simulated using a low value of $a$ ($\leq$ 0.05 \AA$^{-3}$). For such low $a$ values and 0.2/0.6M solutions, the time $t_2^*$ may be as large as 1.3-1.8 ns, a value that is one order of magnitude larger than its \allatom counterpart. In that case we note also the near total absence of clusters whose size is larger than 2.

\begin{figure}
\includegraphics[scale=.65]{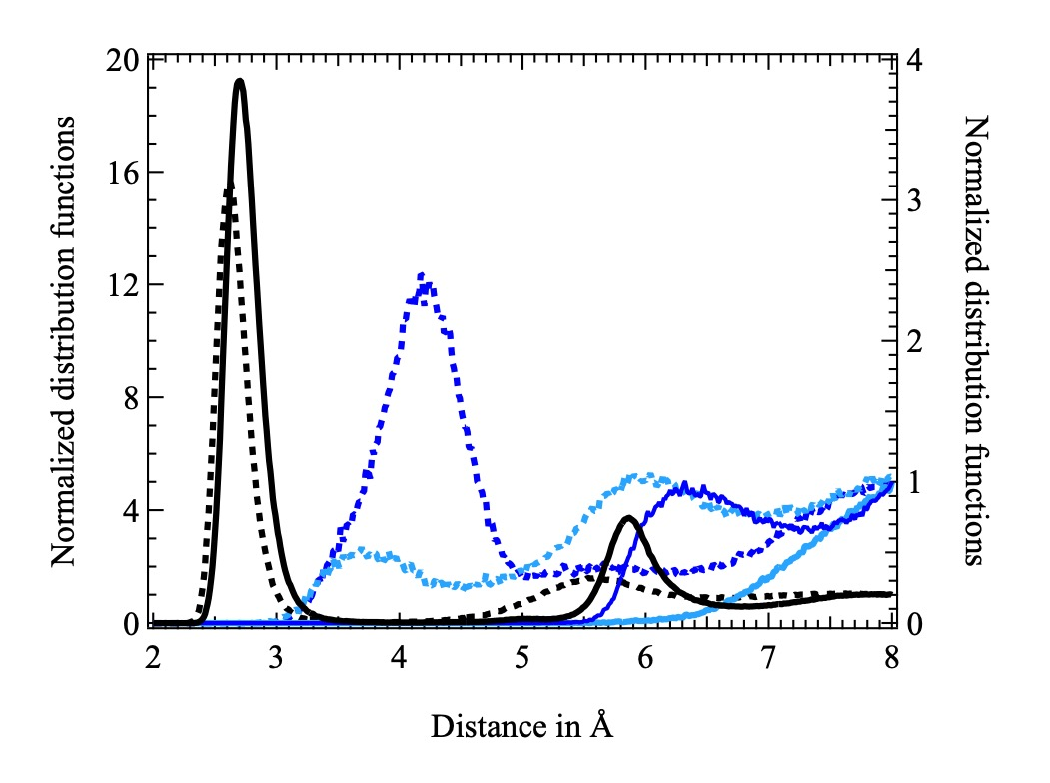}
 \caption{Ion/ion radial distribution functions for a 0.6M NaCl solution in a neat water. Black, blue and light blue: [\na,\cl] (left axis), [\cl,\cl] and [\na,\na] (right axis) distribution functions, that are all normalized to their raw value at 8 \AA, from a simulation along which the damping parameter $a$ is set to 0.05 \AA$^{-3}$. Dashed lines: \allatom functions from Ref. \cite{vallet22}.} 
 \label{fig:gdi_clustering}
\end{figure}

 In Figure \ref{fig:gdi_clustering} we plot the ion/ion radial distribution functions computed from the \ppp 0.6M simulations using a damping parameter $a$ set to 0.05 \AA$^{-3}$ and we compare them to our earlier \allatom data (that agree with all the available \allatom studies \cite{vallet22}). The [\na,\cl] function is in line with the \allatom one (positions and heights of its main maxima and minima), whereas we note the shortest [\na,\na] and [\cl,\cl] distances  to be shifted to larger distances by about 2 \AA $ $ as compared to \allatom simulations. We get the same result for 0.2M and 1.0M NaCl solutions. From visual inspection of the MD trajectories, the associated [\na,\cl] pairs in a \pppl fluid appear to be more uniformly distributed in the solvent than along \allatom simulations (along such simulations, the associated pairs seem to form locally large and weakly bonded super clusters). The largest mean distance among [\na,\cl] pairs in a \pppl fluid explains the largest ion pair mean survival time in a \pppl fluid as using a low damping parameter $a$ value as compared to \allatom simulations.

 \begin{table}[htp]
 \small
\caption{\small Ion association in NaCl \pppl solutions as a function of the damping parameter $a$. Free ions: percentage of \na and \cl ions not involved in a cluster. Clusters: occurrence of Na$_x$Cl$_y$ clusters  (whose size $x+y =$ 2, 3 and 4) identified along the simulations and scaled by the number of sampled simulation snapshots (in parentheses, the cluster mean survival time, in ps). Largest cluster: size and occurrence of the largest cluster identified along the simulations. \emph{na}: the statistical data set is too small to compute meaningful averages. (a) and (b): \allatom data from our earlier study \cite{vallet22} dealing about of the 0.2M and 0.6M NaCl aqueous droplets that comprise about 100 k water molecules and extrapolated to the bulk limit, respectively. (c): \allatom data regarding 1.0M NaCl aqueous solutions from Ref. \cite{soniat16}.} 
\label{tab:clusters}
\begin{center}
\begin{tabular}{ccccccc}
\hline
\hline 

$a$ (in \AA$^{-3}$) & \multicolumn{2}{c}{Free ion in \%} &  \multicolumn{3}{c}{Clusters} & Largest cluster \\
                                &      \na            &   \cl              &  2 & 3 & 4 \\ 
 \hline 
\hline 
0.2M \\
 0.30 &  1.47 $\pm$ 8.03 & 31.33 $\pm$ 7.97 &  0.1 (12)  & $< 0.1$  (294) & 0.2 (58) & 101 (1)     \\
 0.20 & 1.86 $\pm$ 8.49 & 33.57 $\pm$ 8.35 &  0.1 (16)  & $< 0.1$ (471) & $< 0.1$ (\emph{na}) & 118 (1)     \\
 0.15 & 26.90 $\pm$ 7.91 & 61.31 $\pm$ 4.28 &  0.3 (54)  & 0.3 (1058) & 0.1 (85) & 28 (3)     \\
 0.10 & 80.24 $\pm$ 4.43 & 87.12 $\pm$ 2.77 &  0.5 (122)  & 0.2 (326) & $< 0.1$ (\emph{na}) & 5 (54)     \\ 
 0.08 &  92.09 $\pm$ 2.72 & 92.72 $\pm$ 2.43 & 0.3 (248)  & 0.1 (99) & $< 0.1$ (\emph{na})  &  5 (2) \\
 0.05 & 95.35 $\pm$ 1.90 & 95.34 $\pm$ 1.90 &  0.1 (541)  & 0.0 (\emph{na}) & 0.0   (\emph{na}) & 3 (36) \\
 0.03 & 94.08 $\pm$ 2.42 & 94.08 $\pm$ 2.42 & 0.1 (1342) & 0.0(\emph{na})  & 0.0  (\emph{na})  & 3 (4) \\
\allatom$^{(a)}$ & 91.3 $\pm$ 2.0 & 90.6 $\pm$ 2.0 & 6.2 (130) & 0.8 (187)  & 0.2 (326) & 6 (3) \\
\allatom$^{(b)}$ & 92.0 $\pm$ 2.0 & 91.8 $\pm$ 2.0 &                 &                 &                & \\
\hline
\hline
0.6M \\
 0.30 &  0.56 $\pm$ 6.49 & 22.66 $\pm$ 6.93 &  0.3 (7)  & 0.1 (125) & 0.2 (68) & 554 (1)     \\
 0.20 & 0.70 $\pm$ 6.77 & 28.2 $\pm$ 5.56 &  0.3 (8)  & 0.1 (124) & 0.2 (125) & 547 (1)     \\
 0.15 & 4.45 $\pm$ 7.21 & 33.04 $\pm$ 6.26 &  0.2 (69)  & 0.3 (383) & 0.2 (101) & 526 (1)     \\
 0.10 & 41.52  $\pm$ 3.42 & 62.92 $\pm$ 3.42 &  3.6 (113)  & 2.5 (295) & 0.3 (29) & 9 (2)     \\ 
 0.08 &  66.68 $\pm$ 3.59 & 71.81 $\pm$ 3.07 & 3.7 (221)  & 2.7 (300) & 0.2 (28)  &  9 (2) \\
 0.05 & 80.64 $\pm$ 2.50 & 80.61 $\pm$ 2.50 &  1.2  (644)  & 0.1 (12) & 0.0   (\emph{na}) & 4 (23) \\
 0.03 & 80.01 $\pm$ 2.82 & 80.01 $\pm$ 2.82 & 0.5 (1814) & 0.0 (\emph{na})  & 0 (\emph{na})  & 3 (78) \\
\allatom$^{(a)}$ & 81.8 $\pm$ 2.0 & 77.5 $\pm$ 2.0 & 6.2 (130) & 0.8 (187)  & 0.2 (326) & 10 (1)\\
\allatom$^{(b)}$ & 85.0 $\pm$ 2.0 & 82.0 $\pm$ 2.0 &  &   &  & \\
\hline
\hline
1.0M \\
 0.30 &  0.02 $\pm$ 0.07 & 24.75 $\pm$ 4.06 &  0.1 (62)  & 0.1 (239) & 0.1 (441) & 849 (1)     \\
 0.20 & 2.24 $\pm$ 1.08 & 7.67 $\pm$ 6.36 &  0.1 (147)  & 0.1 (208) & 0.1 (374) & 944 (1)     \\
 0.15 & 5.56 $\pm$ 0.95 & 13.97 $\pm$ 13.49 &  0.4 (100)  & 0.4 (223) & 0.4 (165) & 861 (1)     \\
 0.10 & 41.32 $\pm$ 2.26 & 61.57 $\pm$ 1.41 &  6.7 (111)  & 5.4 (235) & 0.6 (40) & 9 (6)     \\ 
 0.08 &  66.41 $\pm$ 1.96 & 70.21 $\pm$ 1.59 & 6.7 (201)  & 3.4 (82) & 0.2 (18)  &  6 (14) \\
 0.05 & 80.09 $\pm$ 1.52 & 80.08 $\pm$ 1.52 &  3.8 (303)  & 0.1 (5) & $< 0.1$   (\emph{na}) & 4 (5) \\
 0.03 & 81.51 $\pm$ 1.47 & 81.51 $\pm$ 1.47 & 4.6 (160) & $< 0.1$ (\emph{na})  & 0 (\emph{na})  & 3 (14) \\
\allatom$^{(c)}$ &    84.0    $\pm$ 1.0 &    85.0  $\pm$ 1.0 & 7.0 (-)&  0.7 (-)  & - (-)   & 8 (79) \\
\hline
\hline
\end{tabular}
\end{center}
\label{default}
\end{table}

As the ions are confined in a cubic sub volume of the simulation box and as they do not interact with their periodic images, one may wonder about comparing our \pppl data to earlier \allatom ones in aqueous droplets  and in bulk water. First our earlier results in aqueous droplets show the main features of ion clustering to be almost fully converged to their bulk limit in 100 k water droplets \cite{vallet22}. Moreover ion clustering properties at the air/droplet interface are close to those at the droplet core. The presence of an interface has a priori a weak effect on NaCl association in aqueous finite environments. 

We may also note the solute/solvent polarization interaction energies corresponding to high level \pppl domains to be weak and they rapidly decrease until to be fully negligible for $l = 4$. Regardless of the damping magnitude and the salt concentration, the \pppl/salt polarization energies are all larger than -10 Mcal for $l = 1$ (original \ppp's), whereas they amount at most to about -0.4 and -0.005 Mcal for $ l = 2$ and 3, respectively. That conclusion regarding the overall negligible effect of higher \pppl levels on salt solutions is also supported by 100 ns scale simulations of a 1.0M NaCl solution in a neat \ppp fluid using low damping parameter $a$ values. All the ion clustering properties computed along the latter simulations fully match those corresponding to a four level \pppl fluid. For instance the percentage of free ions along \ppp simulations corresponding to $a =0.05$ \AA$^{-3}$ agrees with \pppl data within less than 1 \% (and that percentage decreases as in \pppl simulations for larger $a$ values).

\begin{figure}
\includegraphics[scale=1.0]{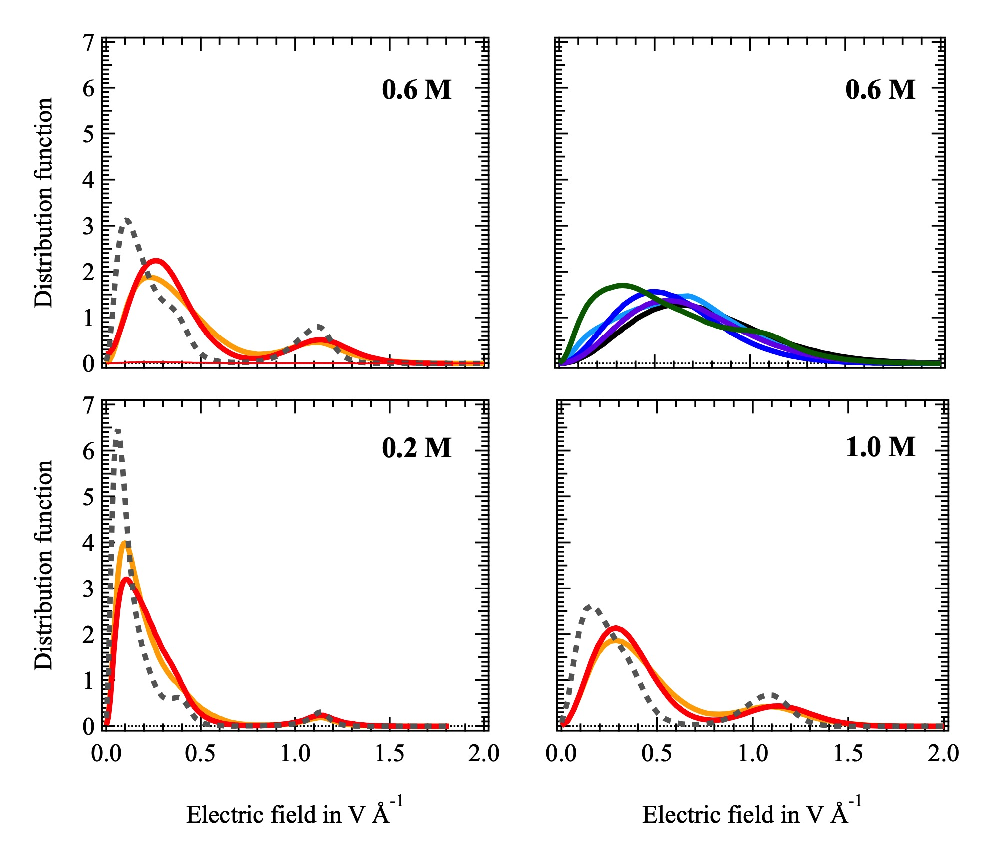}
 \caption{Normalized distribution functions $g(\bar{\mathrm{E}})$ of the total internal electric field on \na ions in NaCl \pppl solutions. Red, orange, green, light blue, blue, violet and black lines: data for a damping parameter $a$ value ranging from 0.03 to 0.30  \AA$^{-3}$, respectively. Dashed lines: normalized distributions of the electric field component arising from ion static electrostatic charges on \na ions and for $a = 0.05$ \AA$^{-3}$. Distribution functions for \cl ions are almost indistinguishable from the present ones.} 
 \label{fig:electric_field}
\end{figure}

We also investigated the mean internal total electric field  acting on \na/\cl ions along our simulations. We plot the normalized distributions $g(\bar{\mathrm{E}})$ of those electric fields for \na ions in Figure \ref{fig:electric_field}. These distributions, which are almost indistinguishable for both ions, strongly depend on the damping parameter $a$. For moderate to large $a$ values, they are mono modal (and they roughly correspond to Gaussian functions), whereas for weakest $a$ values ($\leq 0.05 $ \AA$^{-3}$) they are bimodal, with a main dominant peak located at weak electric field intensities ($< 0.4$ \va) and a smaller peak located at about 1.15 $\pm$ 0.10 \va. The location of the dominant peak increases from 0.15 to 0.30 V \AA$^{-1}$ and its surface  is 10 (0.2M) and 4 (0.6-1.0M) times larger than for the smaller peak. Distributions $g(\bar{\mathrm{E}})$ computed for ions within a sphere of radius 20 \AA $ $ located at the center of the simulation boxes are almost indistinguishable from the latter ones, regardless of $a$. Thus the bimodal $g(\bar{\mathrm{E}})$ distributions for the weakest $a$ values do not arise from an artifact tied to the presence of an interface between the internal salty \ppp domain and the external neat \pppl one in our simulations.

The location of the main distribution $g(\bar{\mathrm{E}})$ peak for the weakest $a$ values ($\leq 0.05 $ \AA$^{-3}$)  for 0.6-1.0M salt concentration are in line with that reported from standard non polarizable \allatom simulations for concentrated NaCl salt aqueous solutions ($\geq$ 1.0M), about 0.50 \va \cite{sellner13}. However \allatom simulations yielded mono modal $g(\bar{\mathrm{E}})$ distributions. The origin of the bimodal $g(\bar{\mathrm{E}})$ distributions in our \pppl simulations using a weak $a$ values arise from the electric field component tied to ion static charges (see Figure \ref{fig:electric_field}). That suggests the structure of molar scale NaCl aqueous solutions predicted using the \pppl approach to not fully match that predicted by \allatom simulations, as already discussed above. However our \pppl approach predicts strong internal electric fields to exist in molar scale NaCl solutions, electric fields whose intensity agrees with that computed from \allatom simulations.
 
\subsection{A hydrophobic polyelectrolyte polymer and at infinite dilution conditions}
 
\begin{figure}
\includegraphics[scale=.75]{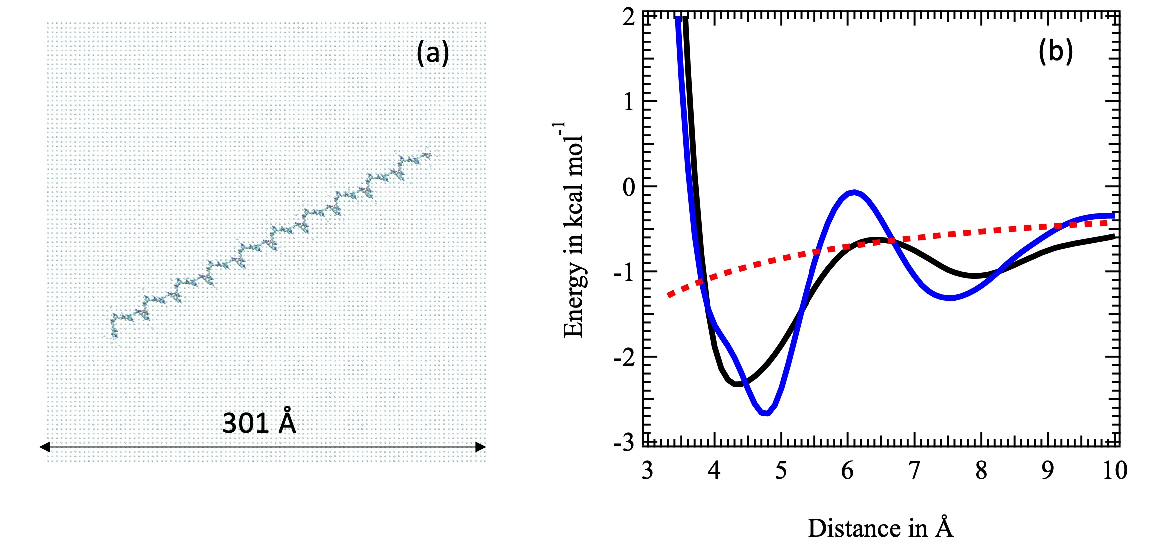}
 \caption{The polymer \hpp and the new \ppp model. (a) : simulation starting structure of  the 10 unit \hpp polymer with its counter ions \cl (in red) embedded in a 0.9M \ppp cubic box. (b) : tetra methyl ammonium/\cl PMF's in a four level \pppl fluid (in blue) and in neat water using a polarizable \allatom approach \cite{houriez19} (in black). Both the latter PMF's are shifted to best reproduce the expected Coulombic potential (dashed red line) within the ion separation distance range 12-16 \AA. Regarding the \allatom PMF, the tetra methyl ammonium/\cl force field parameters are set to reproduce high level quantum \emph{ab initio} data regarding that ion pair in gas phase, as in our recent study dealing about NaCl salt solutions \cite{vallet22}.} 
 \label{fig:polymer}
\end{figure}

In an recent study \cite{masella21}, we discussed simulations dealing about the structural properties of a 10 units hydrophobic polyelectrolyte polymer (denoted as \hpp) in neat \ppp and \pppa fluids at infinite dilution conditions and in presence of counter ions \cl. Each \hpp unit comprises seven di-allyl di-methyl ammonium cations and three acrylamide groups, see Figure \ref{fig:polymer}(a). The polymer total charge is +70 $e$. In that former study, the damping parameters $a$ for all the ionic entities was set to the large value 0.3 \AA$^{-3}$. Starting from a quasi-linear conformation, \hpp rapidly collapses towards a globular form surrounded by a spherical counter ion cloud, a structure that is  stabilized in particular by intra solute polarization forces in both \pppl fluids. 

To discuss the reliability of that result, we built a new \ppp model as detailed above for NaCl in which we systematically set the damping  parameter $a$ for all ionic species to 0.05 \AA$^{-3}$, whereas we consider our earlier \ppp/solute parameter set for the other kinds of polymer moieties. To assess the new \ppp model, we first computed the PMF of the tetra methyl ammonium/chloride anion [\chem{(CH_3)_4N^+}, \cl] ion pair dissolved in a four level \pppl fluid from 10 ns scale simulations. We don't make use here of any correction potential $\delta U^a$. We compare that PMF to that computed in neat water from our own polarizable \allatom approach \cite{houriez19} in Figure \ref{fig:polymer}(b). The new \ppp model is able to overall reproduce the main features of the \allatom PMF. 

We consider that new \ppp model (without any  correction potential $\delta U^a$) to perform two new sets of 10 simulations to investigate the \hpp hydration, using the same computational protocol as in our original study \cite{masella21}. We  thus performed series of simulations of \hpp dissolved in a 0.9M \ppp cubic box using our original simulation starting structure in which the counter ions are located at short range from the \hpp cationic groups. For each simulation the atomic/particle starting velocities are randomly set and \hpp is restrained to its starting linear geometry until an initial relaxation phase of 0.5 ns is achieved. \hpp is dissolved in a \ppp and in a \pppa fluid in the first and second simulation series. We set $R_\mathrm{cut,1}^{pol}$ to 12 \AA. For \pppa simulations, we set $R_\mathrm{cut,2}^{pol}$ to 150 \AA. The \pppa particles undergo the electric field arising from only the static charges of 
ionic species (\hpp cationic groups and counter ions \cl).  

Because of the fast structural evolution of \hpp as interacting with its counter ion cloud along all the new simulations, we stopped them once the \hpp structural transition is achieved, usually within a few ns of simulation. We also accordingly investigate the behavior of \hpp in absence of counter ions and as dissolved in neat \ppp, \pppa and $ppp^3$ fluids from simulations at the 100, 30 and 10 ns scale, respectively. The cut off distances $R_\mathrm{cut,1-3}^{pol}$ are set to 12, 150 and 440 \AA {} and the convergence dipole criterium in all simulations is set as for NaCl solutions. When \hpp  is dissolved in a $ppp^3$ fluid,  we then simulate that polymer as dissolved in an aqueous volume comprising the equivalent of 66 M explicit water molecules. 

\begin{figure}
\includegraphics[scale=.575]{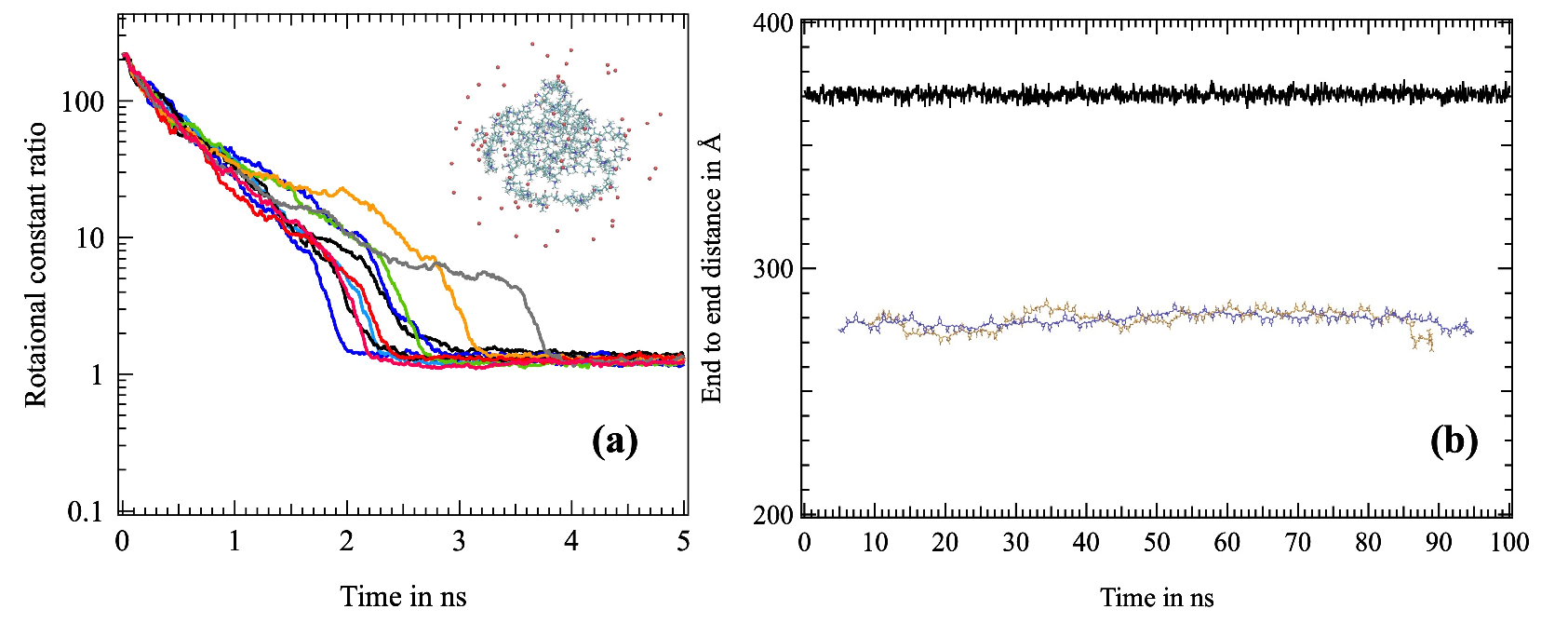}
 \caption{Temporal evolution of the \hpp conformation. The time origin is set to the date at which the harmonic constrains restraining \hpp to its starting linear geometry are removed. (a) : evolution of the ratio between the largest and smallest rotational contants of \hpp in presence of counter ions in a \pppa fluid. The inset shows the final globular structure of \hpp surrounded by a spherical counter ion cloud (red spheres). (b) : temporal evolution of the \hpp  end-to-end distance in absence of counter ions along the \ppp simulation (black line) once a initial relaxation phase of 5 ns is achieved. The inset shows the final quasi linear conformations of \hpp along \ppp and \pppa simulations (in blue and orange, respectively). From the mean distances between two successive carbons interconnecting the diallyl dimethyl ammonium and the acrylamide sub units of \hpp in its quasi linear geometries, we estimate the \hpp contour length to amount to 425 $\pm$ 10 \AA.} 
 \label{fig:polymer_structure}
\end{figure}

In presence of counter ions and as reported in our earlier study \cite{masella21}, \hpp again rapidly evolves towards a globular form (whose radius is about 15 \AA) surrounded by a spherical counter ion cloud along all simulations, see Figure \ref{fig:polymer_structure}(a). That conformational transition is again driven by polarization forces: intra solute Coulombic interactions disfavor the \hpp globular form and its spherical counter ion cloud as compared to the starting linear \hpp structure by about +6.5 \mcal, whereas the system total polarization energy stabilizes the globular form (including \cl) by about -10.0 \mcal. Contrary to NaCl solutions, ion/solvent short range electrostatic damping does not thus appear to play a pivotal role as modeling the hydration of \hpp in presence of counter ions. That may arise from the overall large size of tetra methyl ammonium groups. From a simulation of a tetra methyl ammonium cation dissolved alone in a \ppp box, we estimate the mean average distance between the \ppp's and nitrogen in the cation first hydration shell to be about 4.25 \AA $ $, a distance close to the damping maximum range, 5.5 \AA. For \na and \cl, their first hydration shell radii in a neat \ppp fluid are about 2.3 and 3.3 \AA, respectively.

In absence of counter ions, the \hpp  conformation is stable along the full trajectories in \ppp,  \pppa and $ppp^3$ fluids. It corresponds to an elongated and quasi-linear conformation regardless of the solvent extension, see Figure \ref{fig:polymer_structure}(b). The mean \hpp end to end distance $d_{ee}$ (measured from the first to the last polymer nitrogen) along the simulations is about 360 $\pm$ 10 \AA. We may reasonably assume that magnitude of $d_{ee}$ to be a reliable estimate of the \hpp persistence length. 

To discuss potential drawbacks arising from  the use of a short cut off distance $R_\mathrm{cut,1}^{pol}$ and from taking into account only electrostatic interactions between \hpp charged groups and high level $ppp^{n \geq 2}$ particles, we computed the mean electrostatic energy components $\bar{U}_\mathrm{pol}^{ppp^{1,2}}$ corresponding to the interactions of the full solute (\hpp + counter ions if they are accounted for) with each kind of \ppp and \pppa particles in simulation segments along which the final  polymer conformation is stable. Those $\bar{U}_\mathrm{pol}^{ppp^{1,2}}$ energies are computed for $R_\mathrm{cut,1}^{pol}$ distances that span from 12 to 36 \AA $ $ (whereas $R_\mathrm{cut,2}^{pol}$ is systematically set to 150 \AA). We also computed along those segments the mean intra solute potential energy $\bar{U}^{\hpp}$. For comparison purpose we add to the value $\bar{U}^{\hpp}$ corresponding to \hpp simulated in absence of counter ions, the electrostatic interaction energy $ \bar{U}$(\cl) corresponding to 70  \cl anions, each of them dissolved alone in an infinitively large \ppp fluid. We estimated $ \bar{U}$(\cl) from a simulation of a single \cl anion dissolved in a small \ppp box (which comprises about 1 000 \ppp's and we set $R_\mathrm{cut,1}^{pol}$ to 12 \AA), and we add to the mean anion/\ppp electrostatic energy  the long range correction $\delta G_\mathrm{lr}$ detailed in Section \ref{sec:cg}. That yields $ \bar{U}$(\cl) = -6.1 \mcal. Regarding \hpp in presence of counter ions,  the  energies $\bar{U}_\mathrm{pol}^{ppp^{1,2}}$ are averaged over the final segments of the 10 simulation set. Note that as reported in our original study \cite{masella21}, the \ppp local densities are here also fully converged to the mean bulk value 14 \AA $ $ away from the polymer globular "surface", as well as from the main polymer chain as in the quasi linear conformation.

\begin{figure}
\includegraphics[scale=.65]{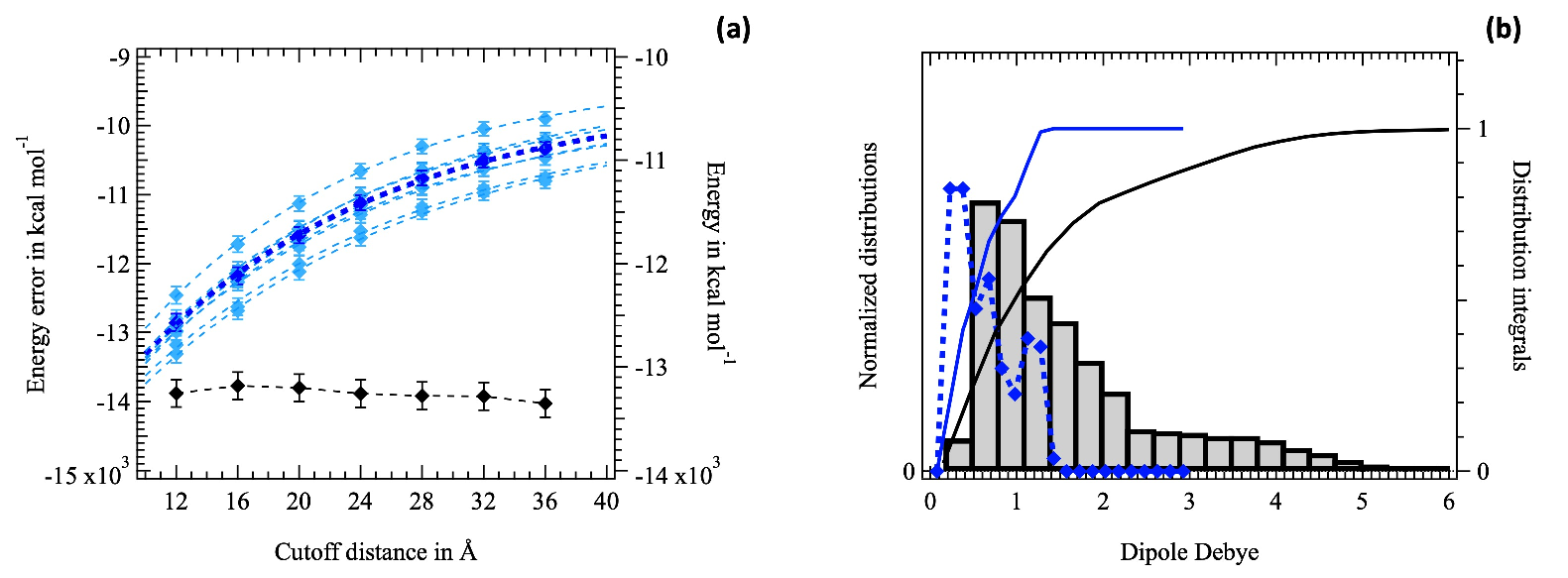}
 \caption{(a) : interaction energies $\bar{U}_\mathrm{pol}^{ppp^{1+2}}$ as a function of the cut off distance  $R_\mathrm{cut,1}^{pol}$. Black symbols (left axis), light and dark blue symbols (right axis) : energy data for the \hpp globular and quasi linear conformations, respectively. The error bars are the root mean square deviations of the averaged data. Dashed blue lines : best exponential functions $U_\infty +  b_0 \exp \left( -c_0 \times R_\mathrm{cut,1}^{pol} \right)$ whose parameters $U_\infty, b_0$ and $c_0$  are adjusted to reproduce the raw energy data. Regarding the energies $\bar{U}_\mathrm{pol}^{ppp^{1+2}}$, their $\bar{U}_\mathrm{pol}^{ppp^{1}}$ and $\bar{U}_\mathrm{pol}^{ppp^{1}}$ components are in a 3:2 ratio for the linear conformation, whereas that ratio varies from 10:1 up to 100:1 for globular geometries. (b) : normalized distributions of the induced dipole moments on polymer alkyl carbons (left axis) and their corresponding integrals (right axis). Blue and black data correspond to the linear and globular forms, respectively.}
 \label{fig:polymer_energy}
\end{figure}

In Figure \ref{fig:polymer_energy}(a) we plot the energies $\bar{U}_\mathrm{pol}^{ppp^{1}}$,  $\bar{U}_\mathrm{pol}^{ppp^{2}}$ and their sums $\bar{U}_\mathrm{pol}^{ppp^{1+2}}$ as a function of $R_\mathrm{cut,1}^{pol}$. Interestingly, the sum  $\bar{U}_\mathrm{pol}^{ppp^{1+2}}$ for \hpp in a linear conformation (and simulated in absence of explicit counter ions) is constant within 0.4 \%, regardless of the $R_\mathrm{cut,1}^{pol}$ value. In that case $\bar{U}_\mathrm{pol}^{ppp^{1+2}}$  amounts to about -13.25 $\pm$ 0.05 \mcal. However that sum for the \hpp globular forms (taking into account the counter ions) noticeably increases as $R_\mathrm{cut,1}^{pol}$ increases, from -12.5 $\pm$ 0.3 \mcal ($R_\mathrm{cut,1}^{pol}$ = 12 \AA) to -10.7 $\pm$ 0.3 \mcal ($R_\mathrm{cut,1}^{pol}$ = 36 \AA). For \hpp in globular forms, the $\bar{U}_\mathrm{pol}^{ppp^{1+2}}$ data can be accurately reproduced using  an exponential function, which yields an extrapolated value for that energy of -9.60 $\pm$ 0.2 \mcal at $R_\mathrm{cut,1}^{pol}$ = 150 \AA. 

The stronger effect of $R_\mathrm{cut,1}^{pol}$ on the solute/\ppp electrostatic interactions for \hpp in a globular form as compared to a quasi linear one is tied to the magnitude of the \hpp atomic induced dipole moments. In globular conformations, and as already shown \cite{masella21}, the \hpp carbon dipole moment increases as the carbon distance to the polymer center of mass increases up to reach values larger than 4 Debye at the \hpp globular surface. In a quasi elongated conformation (with no counter ion), the carbon induced dipoles are weaker and they never exceed 1.4 Debye, see the dipole distributions in Figure \ref{fig:polymer_energy}(b). The large induced dipole moment values for carbon atoms in globular conformations are in line with the strength of the induced dipole moments of water molecules interacting at short range from highly charged cations (3 Debye and above), as reported from both polarizable \allatom and quantum Car-Parinello simulations \cite{guardia09,real12}. Note that for efficiency reasons the \pppa particles do not undergo the electric field arising from the solute induce dipoles in the present study (see above).

Hence whereas the effect of the solvent granularity at medium range from a highly charged solute like \hpp (regardless of the presence or not of counter ions) is weak on its geometry, it has a much more accented impact on the magnitude of the solute/solvent electrostatic interaction energies. From extrapolated values, the solute/solvent interaction energies $\bar{U}_\mathrm{pol}^{ppp^{1+2}}$  are more stable by 4.7 $\pm$ 0.1 \mcal for \hpp in a linear form than in a globular conformation (and surrounded by a spherical counter ion cloud).

The intra solute energies $\bar{U}^{\hpp}$ amount to +8.20 $\pm$ 0.25 \mcal (\hpp globular forms + counter ions) and +20.05 $\pm$ 0.10 \mcal (\hpp in a linear geometry). As discussed above, strong intra solute polarization interactions between the counter ions and \hpp in a globular form stabilize such kind of geometries, as well as the short range non polar solute/solvent mean interaction energies  $\bar{U}_\mathrm{LJ}^{ppp}$, which amount to -4.09 $\pm$ 0.05 and -2.71 $\pm$ 0.02 \mcal, for the globular and linear \hpp forms, respectively. On the other hand, the energy $\bar{U}_\mathrm{pol}^{ppp^{3}}$ is fully negligible for globular forms whereas it amounts to -2.25 $\pm$ 0.05 \mcal for the linear conformation. Taking also into account the energy $ \bar{U}$(\cl) (-6.1 \mcal)  as well as the long range correction $\delta G_\mathrm{lr}$ corresponding to a +70 $e$ point charge interacting with the solvent extending beyond  440 \AA {} from it (-1.85 \mcal), that yields the \hpp linear conformation to be more stable (enthalpically) at infinite dilution conditions than the globular forms (interacting with counter ions at short range) by about 0.6 \mcal based on $\bar{U}_\mathrm{pol}^{ppp^{1+2}}$ data extrapolated at $R_\mathrm{cut,1}^{pol}$ = 150 \AA. Note also that the difference in intra solvent energies between linear and globular geometries is weak, but still in favor of the linear form by about 0.30 $\pm$ 0.05 \mcal.

As the dissociation of a molecular assembly dissolved in a neat water is usually favored by entropic effects, we may thus reasonably conclude the \hpp elongated and quasi-linear form to be more stable than globular ones at infinite dilution conditions as taking into account an infinitely large \pppl environment and as enhancing the precision of the our \pppl approach as computing medium range solute/solvent interactions (\ie by considering $R_{cut,1}^{pol}$ cut off distance at the 30 \AA $ $ scale).

Besides the overall reduced simulation times that can be readily investigated nowadays using standard MD techniques (which prevent an exhaustive exploration of molecular system potential energy surfaces), our new results thus show that simulating a polyelectrolyte polymer with its counter ions to investigate the polymer behavior at infinite dilution conditions is questionable, especially in the case of polyelectrolyte polymers comprising large polarizable and hydrophobic sub units and whose ionic groups are at a long enough distance from each other, like \hpp. That arises in particular from the need of explicit accounting for infinitely large solvent domains, which can not be readily simulated on modern computing systems, even using periodic conditions and standard lattice based numerical schemes.

We may note that simulations in a water droplet using a polarizable \allatom force field \cite{masella21} showed \hpp globular forms surrounded by a spherical counter ion cloud to be stable. Such compact geometries thus correspond at least to meta stable states in aqueous environments, regardless of the description of the solute/solvent interactions. Hence stable polyelectrolyte polymer globular forms surrounded by spherical counter ion clouds as predicted by our multi scale simulation approach for \hpp do not a priori arise from artifacts tied to our solvent approach. Note also that globular forms for hydrophobic polyelectrolyte polymers in presence of counter ions are not systematically predicted by our multi scale \pppl approach. In our earlier study \cite{masella21} we also reported simulations (at the 100 ns scale) of carboxylated polystyrenes with a high fraction charge and in presence of \na counter ions (that are also located close to the polymer anionic heads in the simulation starting structures). Those kinds of polyelectrolyte polymer do not evolve towards a globular form  as simulated in a \pppl fluid in presence of \na counter ions and their linear starting structure is stable along full 100 ns trajectories.

 \section{Conclusion}

 In the present study we discussed potential issues that may alter the reliability of simulations performed using the multi-scale polarizable particle \pppl approach to model the solvent water as investigating  the hydration of complex solutes (like NaCl salty aqueous solutions and the hydrophobic polyelectrolyte polymer \hpp). Among the \pppl parameters that handle the solute/solvent interactions, the magnitudes of two of them, namely the extension of the solvent sub domain SD at  the close vicinity of the solute (domain  in which each \ppp particle corresponds to a single water molecule), and the intensity of \ppp/solute atom short range electrostatic damping, are shown to be pivotal. The extension of sub domain SD is a key parameter to achieve a high accuracy as modeling the hydration of highly charged solutes like \hpp or the association of highly charged ion pairs. Interestingly that issue is tied to the modeling of medium range solute/solvent interactions (\ie interactions between \ppp's and solute atoms that lie between 10 to 20 \AA $ $ from each other). However that issue appears to have a weak effect on the structural behavior of a complex polyelectrolyte polymer like \hpp dissolved in a \pppl fluid. For efficiency reason we thus recommend to simulate the hydration of complex solutes using a moderately extended \ppp sub domain SD (\ie a \ppp shell extending up to 12 \AA $ $ from any solute atom) and to post process the simulations using a larger extension for SD to better assess the enthalpic stability of the solute conformations observed along a simulation.

Solvent/solute short range electrostatic damping appears to be pivotal as modeling small ions (like \na and \cl). In order to reproduce data from polarizable \allatom simulations regarding NaCl aqueous solutions at molar/sub molar scale concentrations (data that agree with experiment), we show that solute/solvent short range electrostatic damping effects have to be strong enough to yield a close weight for the electrostatic and non electrostatic components  of the hydration Gibbs free energies of small ions. However for large ions like tetra methyl ammonium, damping effects appear to have a marginal role in modeling the hydration of a complex solute like  \hpp. In all we recommend to systematically consider \ppp/solute strong short range damping to model ion hydration in \pppl fluids. 

Another issue that can affect the conclusions drawn from the \pppl approach, issue that is not specific to it as it can also impact the conclusions drawn from standard \allatom simulations, is the explicit accounting of counter ions to model the hydration of a solute at infinite dilution conditions. For instance simulations of \hpp in presence of counter ions (that are at contact of the \hpp cationic groups in the starting simulation structure) can lead to compact/globular polymer conformations interacting with an external spherical counter ion cloud, assembly that is stabilized by polarization forces. Such a globular conformation, that is also a priori stable as simulated in a water droplet using a polarizable \allatom force field, is shown however to be less enthalpically stable than a \hpp quasi linear conformation fully dissociated from its counter ions by enhancing the precision of the multi scale \pppl approach (that is achieved by enlarging the upper bound of the sub domain SD away from the solute) and by accounting for the long range solute/solvent interactions not accounted for during the simulations and arising from an infinitely large water environment.  We thus recommend to simulate polyelectrolyte polymers, like \hpp,  dissolved alone in water environments in absence of explicit counter ions to investigate their behavior at infinite dilution conditions. However counter intuitive meta stable conformations, like \hpp globular forms, which may be observed as simulating a solute in presence of its counter ions, are also pivotal as building coarse grained approaches from bottom up schemes in order to account for the solute flexibility.

\section*{Acknowledgments}

This work was granted access to the TGCC HPC resources under the Grand Challenge allocation [GC0429] made by GENCI.

\bibliography{bibliography}

\end{document}